\documentclass[]{emulateapj}
\usepackage{graphicx}
\usepackage{amssymb}
\usepackage{amsmath}
\usepackage{wasysym}

\shorttitle{Solar system oxygen}
\shortauthors{Young et al.}

\begin{document}

\title{Astronomical Oxygen Isotopic Evidence for Supernova Enrichment of the Solar
System Birth Environment by Propagating Star Formation}

\author{Edward D. Young\altaffilmark{1,2}, Matthieu Gounelle\altaffilmark{3}, Rachel L. Smith\altaffilmark{4},
Mark R. Morris\altaffilmark{5}, Klaus M. Pontoppidan\altaffilmark{6}}

\altaffiltext{1} {Department of Earth and Space Sciences and Institute of Geophysics and Planetary Physics, University of California Los Angeles, 595 Charles E. Young Drive East, 
Geology Building, Los Angeles, CA 90095-1567, USA; eyoung@ess.ucla.edu}
\altaffiltext{2} {Institute of Geophysics and Planetary Physics, University of California Los Angeles}
\altaffiltext{3} {Laboratoire  de Min\'{e}ralogie et de Cosmochimie
du Mus\'{e}um, CNRS \& Mus\'{e}um National d'Histoire Naturelle, UMR
7202, CP52, 57 rue Cuvier, 75 005 Paris, France; gounelle@mnhn.fr}
\altaffiltext{4} {Department of Earth and Space Sciences, University of California Los Angeles, 595 Charles E. Young Drive East, Geology Building, Los Angeles, CA 90095-1567, USA; rsmith@ess.ucla.edu}
\altaffiltext{5} {Division of Astronomy and Astrophysics, Department of Physics and Astronomy, University of California, Los Angeles,Physics and Astronomy Building, 430 Portola Plaza, Los Angeles, CA 90095, USA; morris@astro.ucla.edu}
\altaffiltext{6} {Division of Geological and Planetary Sciences, California Institute of Technology, Pasadena, CA 91125; pontoppi@gps.caltech.edu.}

\begin{abstract}
New infrared absorption measurements of oxygen isotope ratios in CO gas from individual young stellar objects confirm that the solar system is anomalously high in its $\rm [^{18}O]/[^{17}O]$  ratio compared with extra-solar oxygen in the Galaxy.  We show that this difference in oxygen isotope ratios is best explained by $\sim 1$\% enrichment of the proto-solar molecular cloud by ejecta from type II supernovae from a cluster having of order a few hundred stars that predated the Sun by at least 10 to 20 Myr.  The likely source of exogenous oxygen was the explosion of one or more B stars during a process of propagating star formation.
\end{abstract}

\keywords{Galaxy: abundances --- infrared: stars --- radio lines: ism --- Sun: abundances}

\section{INTRODUCTION}

\begin{figure}
\centering
\includegraphics[scale=0.06]{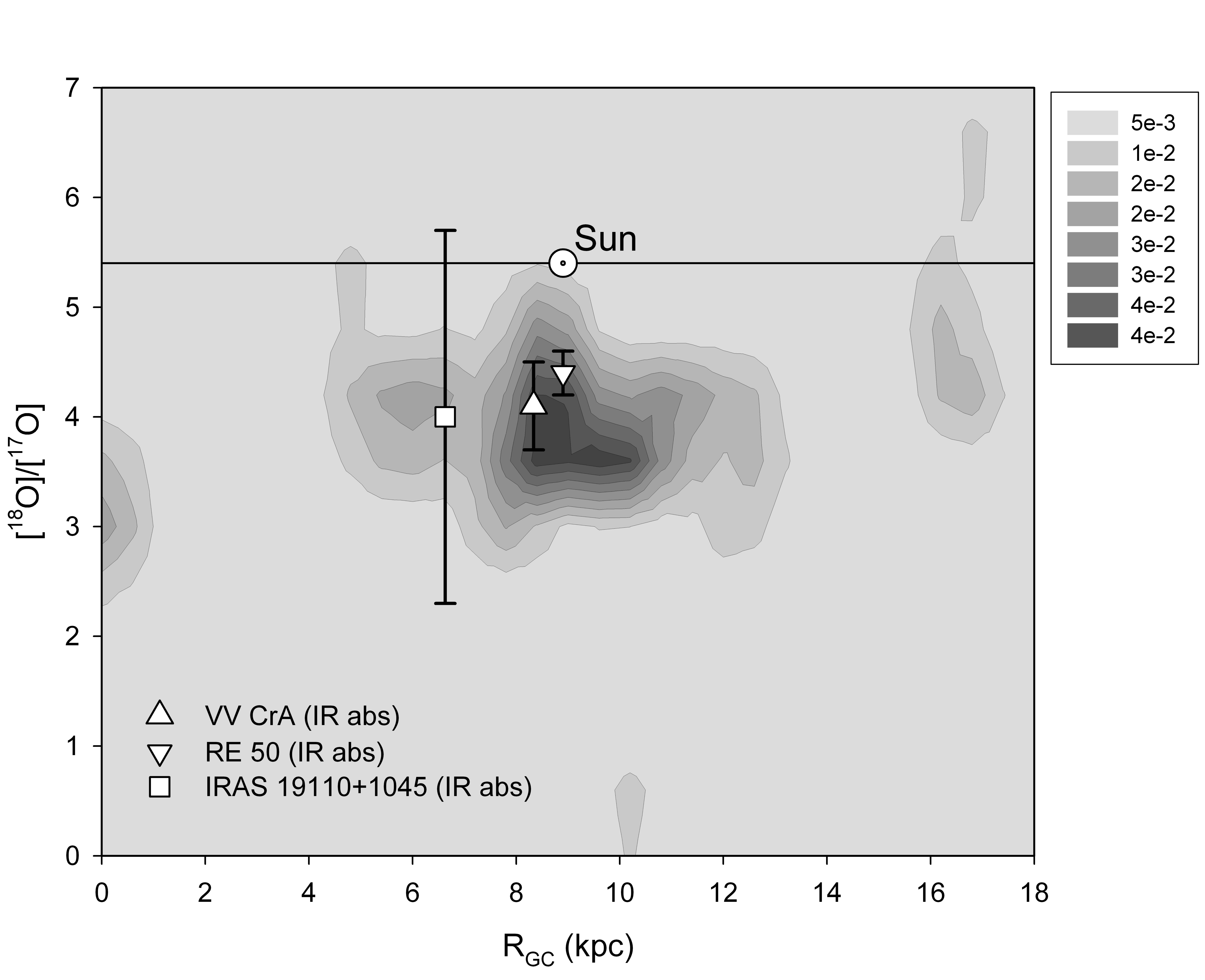}
\caption{Plot of $\rm [C^{18}O]/[C^{17}O]$ vs. distance from the Galactic center ($\rm R_{GC}$) for previously published molecular cloud data \citep{Penzias1981-518,Wouterloot2008-237} (contours), the solar system, and the new survey of young stellar objects (data points with 2$\sigma$ error bars).  The intensity of grey scale from light grey to black shows the probability density of previous molecular cloud $\rm [C^{18}O]/[C^{17}O]$ measurements. }
\label{radio_data}
\end{figure}

Solar system oxygen isotope ratios are peculiar with respect to Galactic values.  All solar system materials have $\rm [^{18}O]/[^{17}O]$ ratios of 5.2 $\pm$ 0.2 ($\rm [^{18}O]$ refers to the abundance of $\rm ^{18}O$ by number) (see Appendix). Values from measurements of radio emission from isotopologues of CO, OH,  $\rm H_2CO$ and $\rm HCO^+$  in molecular clouds across the Galaxy, on the other hand, are between $3.5 \pm 0.3$ \citep{Wannier1989-107,Penzias1981-518} and $4.1 \pm 0.1$ \citep{Wouterloot2008-237}.  Figure \ref{radio_data} shows radio emission data for CO oxygen isotopologues in molecular clouds spanning a large range of distances from the Galactic center.  Both the older data \citep{Penzias1981-518} and the revised data \citep{Wouterloot2008-237} are included.  The latter differ from the former by inclusion of more rotational transitions and a model for optical depth effects.  Also shown is the $\rm [^{18}O]/[^{17}O]$ for the solar system.  The radio data are shown together as probability density contours that are essentially a smoothed two-dimensional histogram.  The method for contouring the data is described in \S 4.2.  It is clear from Figure \ref{radio_data} that the solar ratio of the rare oxygen isotopes is inconsistent with the vast majority of Galactic values.  The newer data also suggest a hint of a trend towards greater $\rm ^{17}O$ relative to $\rm ^{18}O$ in the Galactic center and an excess of $\rm ^{18}O$ relative to $\rm ^{17}O$ in the outer-most Galaxy.  We note that in view of radial rates of stellar migration in the Galaxy, the Sun could not have formed near the Galactic center nor could it have formed beyond $\sim 10$ kpc from the Galactic center \citep[e.g., ][]{Wielen1996-314}.

Explanations for this disparity in solar and Galactic oxygen isotope ratios have included: 1) systematic errors between radio emission data used to measure molecular cloud $\rm [^{18}O]/[^{17}O]$ and mass spectrometry methods used to measure solar values \citep{Prantzos1996-760}; 2) a burst of high-mass star formation produced by merger of another galaxy with the Milky Way approximately 5 to 6 Gyr before present \citep{Clayton2003-313}; and 3) Galactic chemical evolution of $\rm [^{18}O]/[^{17}O]$ over the past few billion years \citep{Nittler2009-271}.   In this paper we suggest that the peculiar $\rm [^{18}O]/[^{17}O]$ of the solar system is a consequence of local enrichment of the solar birth environment by explosion of nearby B stars.  The work is motivated by our recent infrared absorption data for young stellar objects \citep{Smith2009-163} showing that systematic errors are not likely to be the cause of the disparity.  We show that the oxygen isotopic composition of the exogenous type II supernova ejecta required to explain the solar ratio of rare oxygen isotopes  places constraints on the maximum masses of the supernova progenitors.  The limit on progenitor masses in turn suggests a scenario of enrichment by propagating star formation leading up to the formation of the Sun.  This  scenario is distinct from previous suggestions of supernova enrichment in which it had been assumed that the supernova progenitors were coeval with the  Sun \citep[e.g., ][]{Schramm1982-236}.  It is consistent, however, with a proposed origin for short-lived radionuclides in the early solar system in which several supernovae from a previous generation of star formation enrich nearby molecular cloud material prior to collapse to form a new generation of  stars \citep{Gounelle2009-1}.  

The isotopic compositions of oxygen are reported here in plots of $\rm [^{17}O]/[^{16}O]$ against $\rm [^{18}O]/[^{16}O]$, referred to commonly as ``three-isotope'' plots, throughout this paper.  Because the differences in isotope ratios we are concerned with are large, we will use a variant on the delta notation commonly used in the cosmochemistry literature to report isotope ratios.  In this notation, ${\rm \delta^{17}O'} = 10^3 \ln ({\rm [^{17}O]/[^{16}O]}_i/{\rm [^{17}O]/[^{16}O]}_{\rm ISM})$ where ISM refers to the local ISM as defined in \cite{Wilson1999-143} (an arbitrary but convenient reference with $\rm [^{16}O]/[^{18}O] = 557$ and $\rm [^{16}O]/[^{17}O] = 2005.2$) and $i$ refers to an object or mass interval of interest.  Positive values for $\rm \delta^{17}O'$ signify enrichment in $\rm ^{17}O$ relative to $\rm ^{16}O$ with respect to the local ISM while negative values for $\rm \delta^{17}O'$ signify a relative depletion.  Values for $\rm \delta^{18}O'$ are defined analogously (the prime in each instance signifies this logarithmic form of the delta notation).

\section{GALACTIC CHEMICAL EVOLUTION OF  $\rm ^{18}O/^{17}O$ }
Both $\rm ^{17}O$ and $\rm ^{18}O$ are secondary nuclides, produced by H and He burning, respectively \citep{Meyer2008-31}.  One expects their solar ratio to be that of the bulk of the Galaxy at the time  of solar system formation unless the abundances of the oxygen isotopes were affected by some local perturbation. Quantitative models for the increase in the abundances of the oxygen isotopes with time as part of the overall Galactic chemical evolution (GCE) show that we should expect that even as $\rm [^{17}O]/[^{16}O]$ and $\rm [^{18}O]/[^{16}O]$ rose linearly with time, the ratio of the two secondary nuclides, $\rm [^{18}O]/[^{17}O]$, should have been nearly constant after the first billion years \citep{Woosley1995-181,Prantzos1996-760}.  A constant $\rm [^{18}O]/[^{17}O]$ over time would mean that the anomalous value for the solar system compared with the present-day Galaxy cannot be attributed to GCE over the past 4.6 billion years (the age of the solar system).

\begin{figure}
\centering
\includegraphics[scale=0.07]{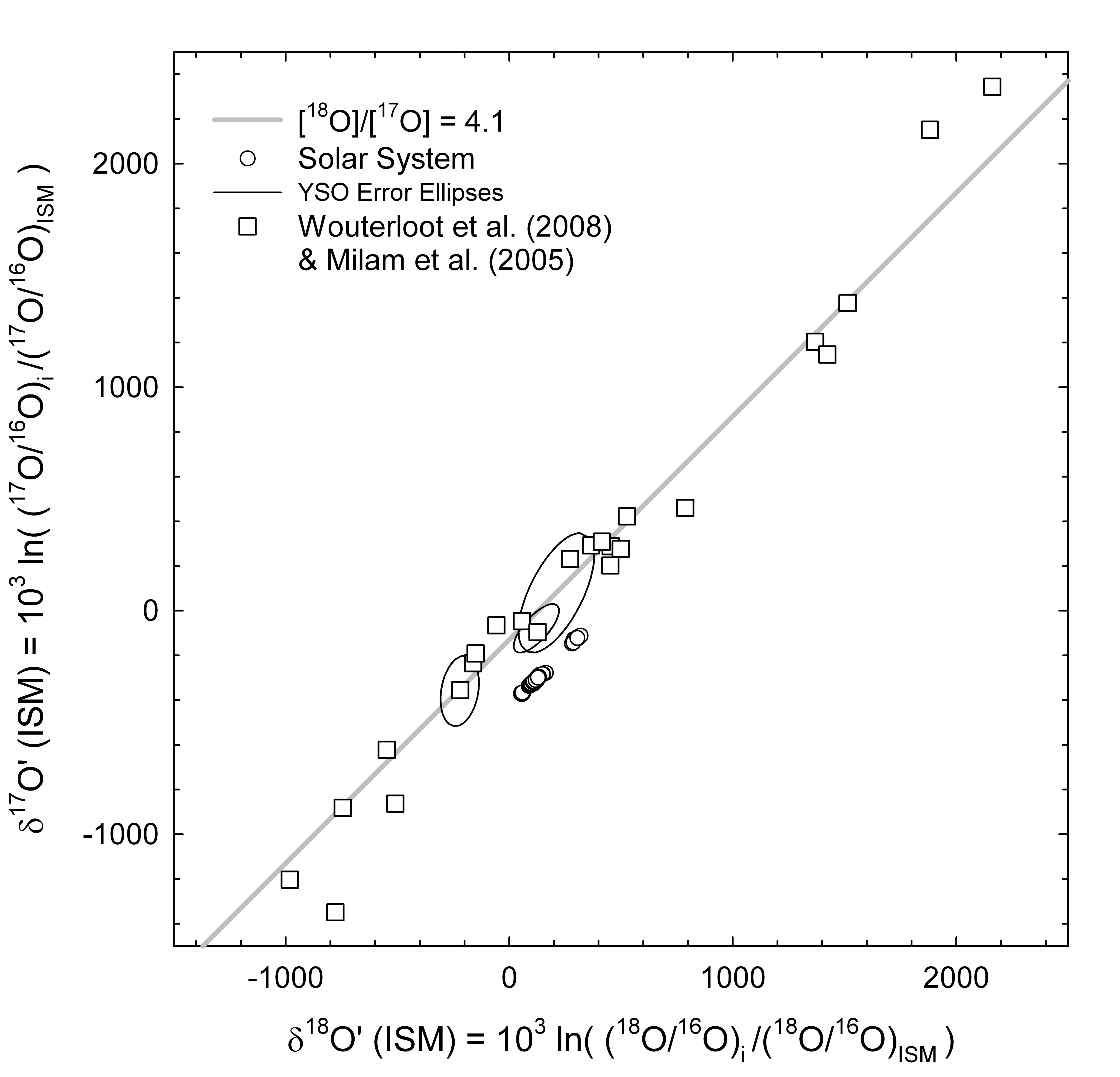}
\caption{Plot of $\rm [^{18}O]/[^{16}O]$ vs. $\rm [^{17}O]/[^{16}O]$ for molecular clouds across the Galaxy in delta notation.  Molecular cloud data (squares) represent $\rm ^{13}CO$, $\rm C^{18}O$ and $\rm C^{17}O$ abundances from Wouterloot et al. (2008) combined with best-fit $\rm ^{12}CO/^{13}CO$ vs. $\rm R_{GC}$ data from Milam et al. (2005).  The effect of using the $\rm [^{12}C]/[^{13}C]$ vs. $\rm R_{GC}$  data for $\rm H_2CO$ and CN rather than CO presented by Milam et al. (2005)  is to move the molecular cloud points up and down the slope-1 line by approximately 200 per mil, but the slope-1 line is preserved.  A best-fit line to the molecular cloud data with equal weighting yields a slope of $1.11 \pm 0.04$ where the 1$\sigma$ uncertainty applies assuming $\pm 100$ for absissca and ordinate for each datum (consistent with a reduced $\chi^2$ of unity).  Also shown are measurements for the solar system compiled from the literature (circles) and 95\% confidence error ellipses for the young stellar objects from Smith et al. (2009) also shown in Figure \ref{radio_data}. }
\label{GCE_3isotope}
\end{figure}

\begin{figure}
\centering
\includegraphics[scale=0.07]{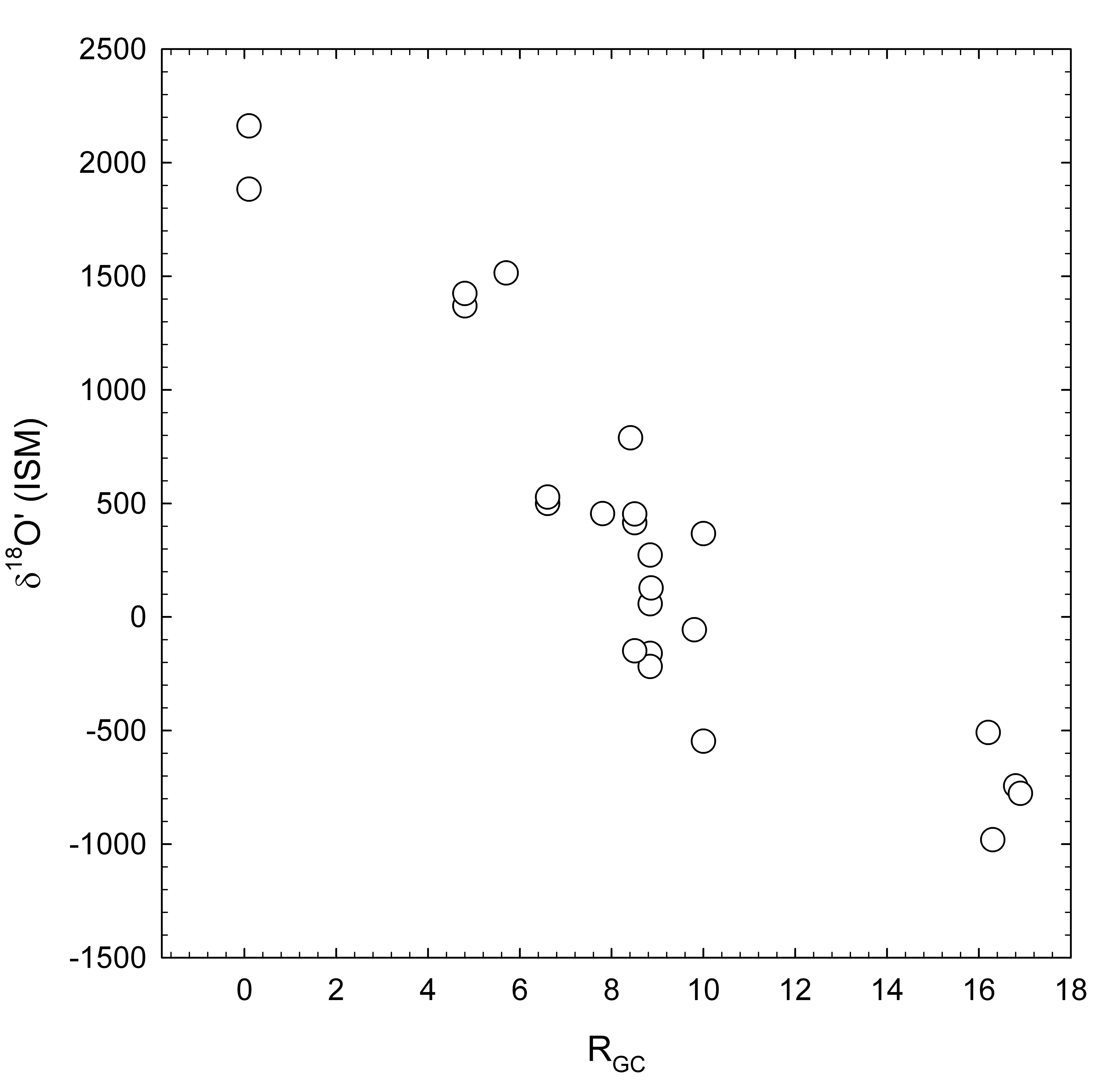}
\caption{Plot of $\rm [C^{18}O]/[C^{16}O]$ vs. distance from the Galactic center ($\rm R_{GC}$) expressed in delta notation relative to the local ISM.   Data represent $\rm ^{13}CO$ and $\rm C^{18}O$ abundances from Wouterloot et al. (2008) combined with best-fit $\rm ^{12}CO/^{13}CO$ data from Milam et al. (2005).  }
\label{GCE_18O}
\end{figure}

A trend of decreasing  $\rm [^{18}O]/[^{16}O]$  and $\rm [^{17}O]/[^{16}O]$ ratios at nearly constant $\rm [^{18}O]/[^{17}O]$ with increasing galactocentric radius, $\rm R_{GC}$ , was well established by early radio emission data \citep{Wilson1999-143}, and the most recent data confirm this trend (Figures  \ref{GCE_3isotope} and \ref{GCE_18O}).  These trends of varying oxygen isotope ratios with  $\rm R_{GC}$ are consistent with our expectations from GCE, where $\rm R_{GC}$ is used as a proxy for time (a well-established means of investigating chemical evolution in the Galaxy).  For this reason we conclude that it is unlikely that a shift in interstellar $\rm [^{18}O]/[^{17}O]$ from 5.2 to 4.1 occurred over the last half to third of the lifetime of the Milky Way at plausible locations for the formation of the Sun (e.g., 10  $>\rm R_{GC}>$  2 kpc).  

\cite{Gaidos2009-163} offer an alternative view of Galactic chemical evolution of oxygen isotopes.  In that work the authors present results of a two-box model calculation in which they posit that the contribution of $\rm ^{17}O$ from AGB stars results in a substantial decline in $\rm [^{18}O]/[^{17}O]$ with time in the Galaxy.  Details of the calculations (e.g., equations representing the box model) are not presented, but the authors note that their model predicts differences between ISM and star forming regions that are not observed \citep[e.g.][]{Smith2009-163}.  They also note that their model is unable to reproduce the oxygen abundances of the ISM 4.6 Gyr before present, presumably meaning the solar value.  The \cite{Gaidos2009-163} model provides no explanation for the trends in oxygen isotope ratios shown in Figures \ref{GCE_3isotope} and \ref{GCE_18O}.  Their GCE evolution model has a slope of nearly 3 on an oxygen three-isotope plot rather than the slope of 1 defined by the molecular cloud CO isotopogue data in Figure \ref{GCE_3isotope}.  The calculations were meant to explain the difference between local ISM and solar $\rm [^{18}O]/[^{17}O]$, but other than this discrepancy (the subject of this paper) we are aware of no data representing the Galaxy that define a slope of $\sim 3$ in three-isotope space.  In support of their model the authors suggest that there is no evidence for the slope-1 line in oxygen three-isotope space.  We find this assertion difficult to defend in view of the CO data shown in Figure \ref{GCE_3isotope}.  Their assertion is made with reference to seven measurements of $\rm [^{16}OH]/[^{18}OH]$ by 
\cite{Polehampton2005-957} that show a mean $\rm [^{16}OH]/[^{18}OH]$  of $424 \pm 70$ $1\sigma$ (corresponding to a $\rm \delta^{18}O'$ value of $+285 \pm 168$) and no apparent trend with Galactocentric radius.   This result is at odds with the substantial gradient shown in Figure \ref{GCE_18O} based on the most recent measurements of molecular cloud CO isotopologues.   Reasons for there being a twenty fold increase in $\rm [C^{16}O]/[C^{18}O]$ with 
$\rm R_{GC}$ but nearly constant $\rm [^{16}OH]/[^{18}OH]$ are unclear.  \cite{Polehampton2005-957} point to low number statistics for their OH measurements and lack of constraints on some of the Galactocentric distances used in their work.  In any case, because CO is the dominant gas-phase reservoir of oxygen in molecular clouds \citep[e.g., ][]{Wakelam2010-21}, we expect this molecule to be, arguably, the most robust tracer for oxygen isotope ratios across the Galaxy.  

\section{OXYGEN ISOTOPE RATIO DATA FOR YOUNG STELLAR OBJECTS}
The possibilities for systematic errors in the  radio data for molecular clouds, or biases due to disparate scales of observation for clouds and the solar system,  are addressed by new data for young stellar objects (YSOs).  We use results from an infrared (IR) spectroscopic survey of molecules in young stellar objects \citep{Pontoppidan2008-1323}, part of which aims to measure oxygen isotope ratios with high precision \citep{Pontopiddan2008-1323,Smith2009-163}. These new data comprise IR absorption spectra for CO rather than radio emission and represent scales of observation commensurate with that of the solar system (on the order of hundreds of astronomical units).   By embodying  an entirely different scale of observation and an independent method of measurement these new data should circumvent sampling bias and/or systematic errors that might be present in the radio emission results.  The purpose in this context is to establish the magnitude and scale of oxygen isotope variability in young stars for comparison with the solar system.  Our survey now includes column densities for CO isotopologues of gas surrounding three young stellar objects, including Reipurth 50 (RE 50), VV Corona Australis (VV CrA), and IRAS 19110+1045.  High-resolution 4.7 $\mu$m fundamental and 2.3 $\mu$m overtone ro-vibrational absorption bands for CO were obtained for RE 50 and VV CrA using   Cryogenic High-Resolution Infrared Echelle Spectrograph (CRIRES) on the Very Large Telescope (VLT) at ESO's Paranal Observatory.  Lower spectral resolution data from Near Infrared Echelle Spectrograph (NIRSPEC) on the Keck II telescope were also analyzed for a more massive embedded young stellar object, IRAS 19110+1045.  Details of the data collection and processing for RE 50 and VV CrA are provided by \cite{Smith2009-163} as part of their study of the potential for photochemical effects on the isotopic composition of CO in YSOs.  Those for IRAS 19110+1045 are given by \cite{Smith2007-158}.  RE 50 is an embedded YSO in the Orion star-forming cloud $\sim 470$ pc from the Sun.  It is an FU Ori type object in stage I.  VV CrA is a stage II T-Tauri disk 130 pc from the Sun.   The third object, IRAS 19110+1045, is a more massive embedded YSO at a Galactocentric radius, $\rm R_{GC}$, of 6 kpc. 

The YSO data are compared with  the molecular cloud radio emission data in Figures  \ref{radio_data} and \ref{GCE_3isotope}.  These results are consistent with the molecular cloud radio emission data in showing that typical Galactic $\rm [^{18}O]/[^{17}O]$ is near 4.  The $\rm [C^{18}O]/[C^{17}O]$ values are $4.1 \pm 0.4$ for VV CrA, $4.4 \pm 0.2$ for RE 50, and $4.0 \pm 1.7$ for IRAS 19110+1045.   We conclude from these measurements that the solar system is indeed unusual, and that the difference between solar $\rm [^{18}O]/[^{17}O]$ and typical Galactic $\rm [^{18}O]/[^{17}O]$ cannot be attributed solely to heterogeneity in oxygen isotope ratios on the scale of individual stars.  A systematic error in the molecular cloud data is also now excluded.

\section{SUPERNOVA ENRICHMENT OF THE PROTOSOLAR MOLECULAR CLOUD}
\subsection{The Oxygen Isotopic Compositions of Supernovae}
We find that the most likely explanation for the anomalous $\rm [^{18}O]/[^{17}O]$ of the solar system is that the Sun and planets formed from molecular cloud material enriched in exogenous $\rm ^{18}O$-rich oxygen ejected from stars in a nearby cluster that ended their lives as type II supernovae (SNe II).  Measurements of the relative abundances of all three stable oxygen isotopes place critical constraints on mixing between normal Galactic oxygen and this $\rm ^{18}O$-rich ejecta.  We explore this proposition, and its consequences for the origin of the solar system, below.

With three isotopes of oxygen we can compare the isotopic composition of the ISM at the time the Sun was formed with  different supernova products in order to identify plausible sources of exogenous oxygen.  The oxygen isotopic composition of the local ISM 4.6 Gyr before present can be estimated from ages of between ~13.6 Gyr and 10 Gyr for the Galaxy and the 4.6 Gyr age of the solar system.  As described above, both $\rm [^{18}O]/[^{16}O]$  and $\rm [^{17}O]/[^{16}O]$ are ratios of secondary to primary nuclides that have risen linearly, to first order, with time in the Galaxy at nearly constant $\rm [^{18}O]/[^{17}O]$.  The rise in the abundances of $\rm ^{18}O$ and $\rm ^{17}O$  relative to that of $\rm ^{16}O$ linearly with time leads to the relationship  $\Delta$age/age $\sim$ $\rm (\Delta [^{18}O]/[^{16}O])/([^{18}O]/[^{16}O])$.  We therefore expect both $\rm [^{18}O]/[^{16}O]$ and $\rm [^{17}O]/[^{16}O]$ ratios in the ISM to have risen by between  35\% and 46\% in the past 4.6 billion years.  More complicated models for oxygen GCE are consistent with this estimate \citep{Prantzos1996-760}.  The $\sim 350$ to 460\permil \, increase in both ratios can be subtracted from the composition of present-day ISM to obtain an estimate of the oxygen isotopic composition of the ISM at the time of the formation of the Sun.  The precision of this estimate is limited by an uncertainty of at least $\sim$20\% ($\pm 200$ \permil)
in $\rm [^{18}O]/[^{16}O]$ and $\rm [^{17}O]/[^{16}O]$ for the present-day ISM.   With this uncertainty, the present-day nominal $\rm [^{18}O]/[^{16}O]$ and $\rm [^{17}O]/[^{16}O]$ values of the local ISM \citep{Wilson1999-143} are reasonable upper limits for the values for the ISM 4.6 Gyr ago. 

Previous work has emphasized that $\rm [^{18}O]/[^{17}O]$ of oxygen liberated by the explosion of a SN II varies systematically with the mass of the progenitor star \citep{Gounelle2007-L123}.  One can illustrate the inputs of individual stars to the composition of stellar ejecta from star clusters using mass fractions over small intervals of the initial mass function.  Starting with the initial mass function, yielding the number of stars $N$ of mass $m$

\begin{equation}
\xi (m) = \frac{{dN}}
{{dm}} = \beta m^{ - \alpha } 
\end{equation}

\noindent integration over some mass interval $m_l$ to $m_u$ (lower to upper) yields for number of stars

\begin{equation}
N_{m_l  \to m_u }  = \beta \int_{m_l }^{m_u } {m^{ - \alpha } } dm
\end{equation}

\noindent and for mass

 \begin{equation}
m_{m_l  \to m_u }  = \int_{m_l }^{m_u } {m\;\xi (m)\;dm} .
\end{equation}
 
\noindent The fraction of mass contained in stars of masses $m_l$ to $m_u$ is then

\begin{equation}
X_{m_l  \to m_u }  = \frac{{\int_{m_l }^{m_u } {m\;\xi (m)\;dm} }}
{{\int_{0.08M \odot }^{100M \odot } {m\;\xi (m)\;dm} }}.
\end{equation}

\begin{figure}
\centering
\includegraphics[scale=0.06]{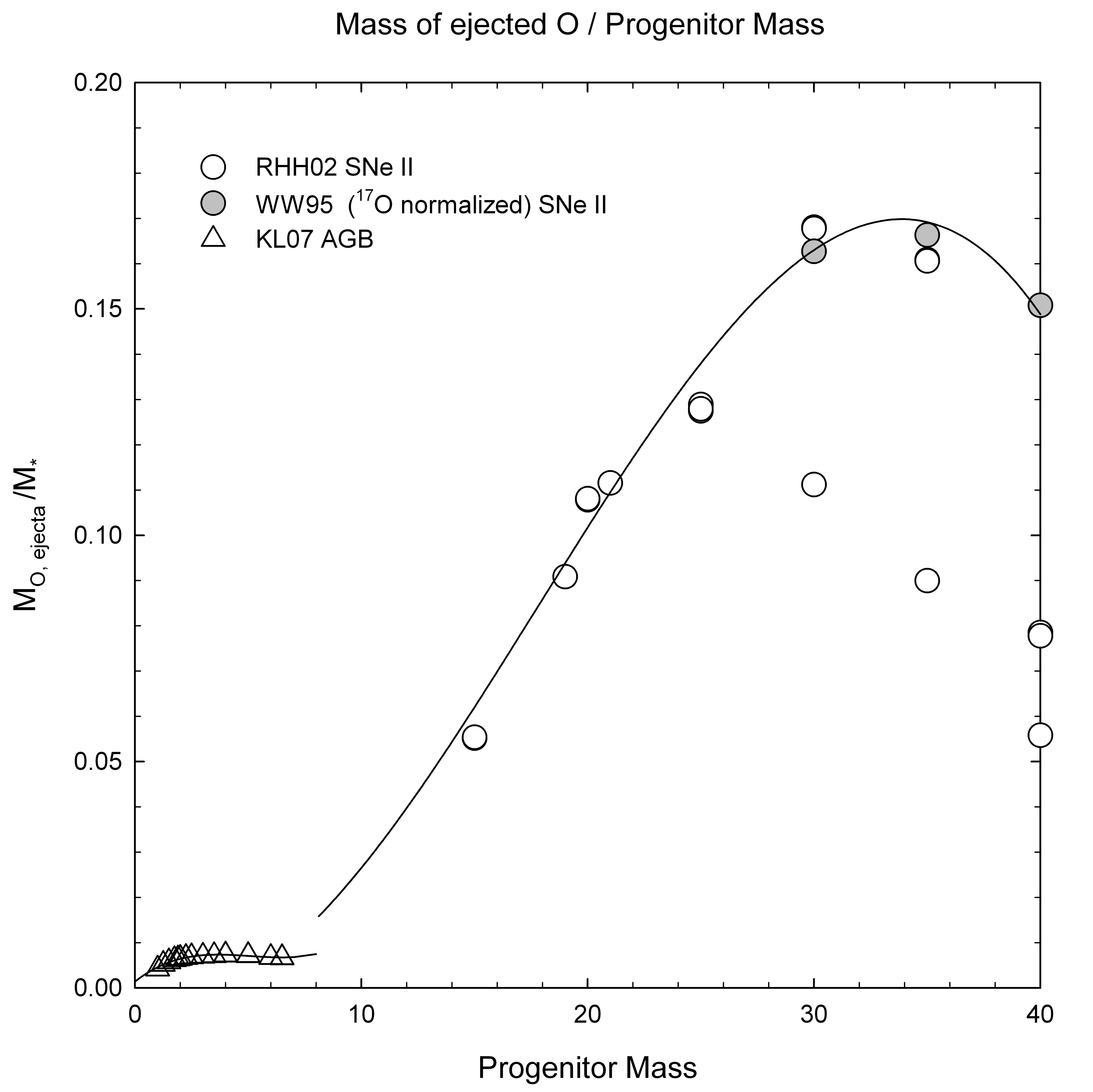}
\caption{Predicted masses of oxygen ejected by type II supernovae as  a function of progenitor mass ($M_{\odot}$) as given by RHHW02 and WW95.  Also shown are the predicted masses of oxygen released by AGB stars as given by KL07.  Fits to these calculations used in the present study are shown as solid lines (Equations (5) and (6)).  Two sets of calculations by RHHW02 are shown based on low and high-energy piston models. }
\label{O_ejecta_yields_bestfit}
\end{figure}

\noindent From these mass fractions of the total stellar system we obtain mass fractions of oxygen ejected from a generation of star formation by fitting existing models for mass loss as functions of progenitor mass. For stars $>$ 8$M_{\odot}$  we used the SNe II yields (by yields we mean production and not net yield that results from considering the nuclides remaining in the stellar core) from \cite{Rauscher2002-323} (RHHW02, including supplemental tables available on line) taking into account also the calculations of \cite{Woosley1995-181} (WW95) and \cite{Woosley2007-269} (WH07, yield and mass cut tables kindly provided by A. Heger).  These studies exhibit large differences in $\rm ^{17}O$ production.  A reduction in $\rm ^{17}O$ yield from WW95 to RHHW02 reflects the revision to the destruction rate for that nuclide \citep{Blackmon95-2642}.  A further reduction in the$\rm ^{17}O$ yield from RHHW02 to WH07 resulted from lower initial CNO abundances of the progenitor stars.  The mass of oxygen ejected from the larger progenitors ($\ge 30 M_{\odot}$) depends on the choice of the intensity of the SN ``piston''  employed by RHHW02.  For the calculations presented here, we made use of the higher-energy piston models.   The resulting yields are similar to those of WW95.  The  oxygen yields as a function of progenitor mass are shown in Figure \ref{O_ejecta_yields_bestfit} along with the fits to the calculations used in the present calculations.   The  fit to the RHHW02 calculations for mass of oxygen ejected relative to mass of progenitor, $M_{\text O}/M_*$, is

\begin{eqnarray}
\frac{{M_{\text{O}} }}
{{M_* }} &=& 1.0250\times 10^{ - 5} M_*^3  + 5.4290\times 10^{ - 4} M_*^2 \nonumber  \\
&-& 1.6617\times 10^{ - 4} M_*  - 1.1800\times 10^{ - 3} 
\end{eqnarray}

 \noindent where masses are in solar  units.  For comparison with the SNe II calculations, and for calculating mass fractions of oxygen relative to all of the oxygen injected into the ISM by a stellar cluster, we include the  oxygen produced by mass loss from AGB stars for $M_* \le 8M_{\odot}$ as calculated by  \cite{Karakas2007-103} (KL07) (Figure \ref{O_ejecta_yields_bestfit}).   The equation for the fit to the KL07 calculations for mass of oxygen released relative to mass of progenitor is
 
 \begin{eqnarray}
\frac{{M_{\text{O}} }}
{{M_* }} &=& 5.5687\times 10^{ - 5} M_*^3  + 8.4985\times 10^{ - 4} M_*^2 \nonumber \\
  &+& 3.9956\times 10^{ - 3} M_*  + 1.3560\times 10^{ - 3} .
\end{eqnarray}
 
\begin{figure}
\centering
\includegraphics[scale=0.07]{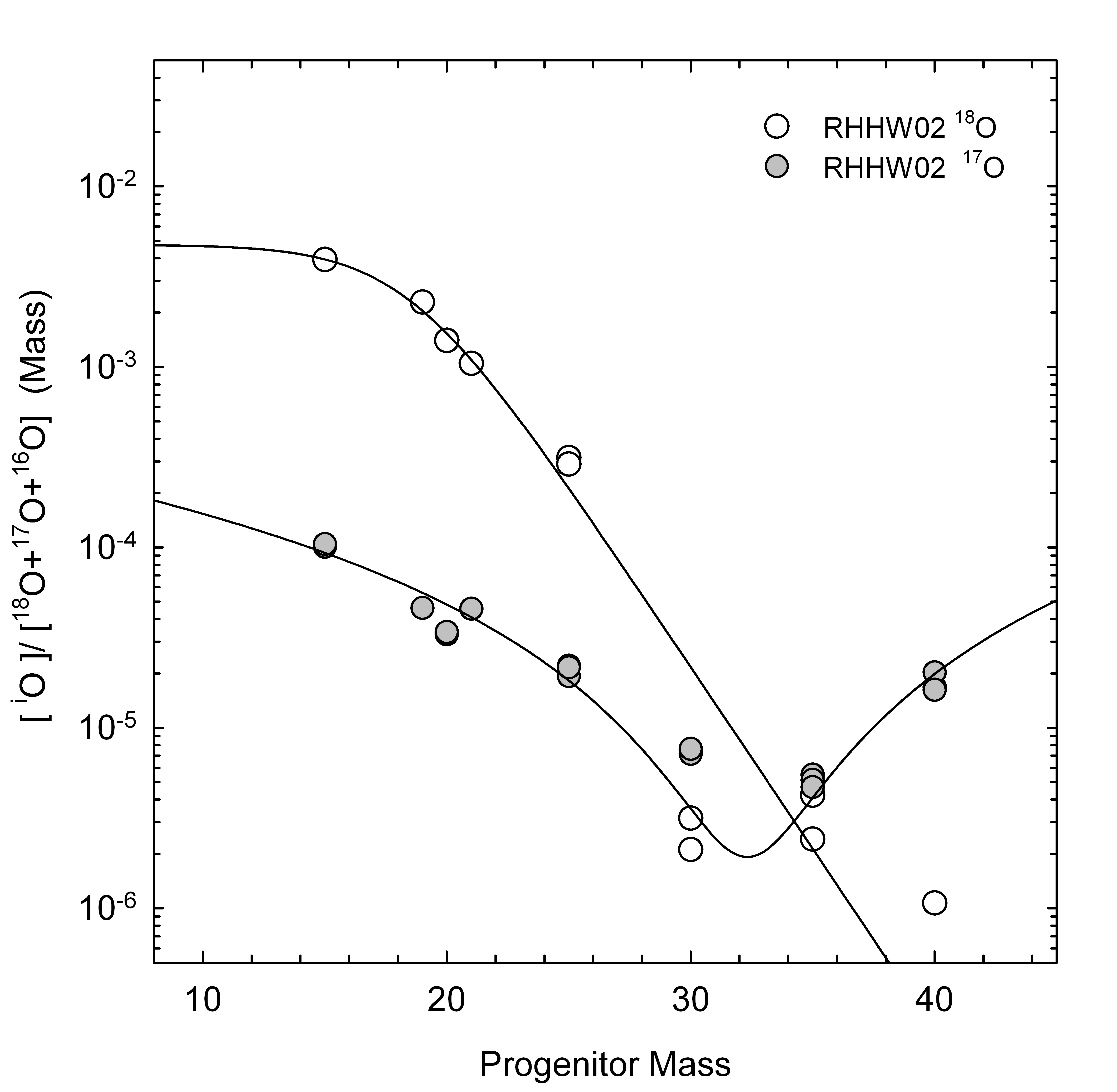}
\caption{Mass fractions of oxygen isotopes $i$, $X^{i}{\text O} = M_{^i{\text O}}/(M_{^{16}{\rm O}}+M_{^{17}{\rm O}}+M_{^{18}{\rm O}})$, as functions of progenitor mass ($M_{\odot}$)as predicted by RHHW02.  Fits used in the present study are shown by the solid curves. } 
\label{isotopes_ejecta_yields}
\end{figure}

 \noindent  From equations (4) through (6) the fraction of oxygen attributable to a progenitor mass $M_*$ in the total ejecta from a generation of stars, $X_{{\text{O,}}\Delta{\text{m}}}$, represented by a specified initial mass function is

 \begin{equation}
X_{{\text{O,}}\Delta {\text{m}}}  = \frac{{\frac{{M_{\text{O}} }}
{{M_* }}X_{\Delta m } }}
{{\sum\limits_{\Delta m} {\frac{{M_O }}
{{M_* }}X_{\Delta m} } }}
\end{equation}
 
\noindent where  $\Delta{\text{m}}$ represents the mass interval $m_l$ to $m_u$, the summation is over all mass intervals, and $M_*$ is the mean mass for that interval.  We are concerned here with both the total oxygen ejected and the isotopic composition of that oxygen.  Oxygen isotope abundances for the ejecta, expressed as the mass fraction of oxygen for nuclide $i$, $X^{i}{\text O} = M_{^i{\text O}}/(M_{^{16}{\rm O}}+M_{^{17}{\rm O}}+M_{^{18}{\rm O}})$, are shown in Figure \ref{isotopes_ejecta_yields}.  The fit to the RHHW02 calculations used here for $\rm ^{18}O$ is

\begin{equation}
X{}^{18}{\text{O  =  }}\frac{{0.0048}}
{{(1 + \exp ( - (M_*  - 18.3973)/( - 2.1545)))}}.
\end{equation}

\noindent For $\rm ^{17}O$ we fit the RHHW02 calculations and then scaled the results by a factor of $0.5$ to bring them into line with the more recent results from WH07.  The fit prior to scaling (Figure \ref{isotopes_ejecta_yields}) is

\begin{eqnarray}
X{}^{17}{\text O} &=  &3.1995\times 10^{ - 4}  - 1.9676\times 10^{ - 5} M_*  \nonumber \\
&+&3.043\times 10^{ - 7} M_*^2. 
\end{eqnarray}

\begin{figure}
\centering
\includegraphics[scale=0.07]{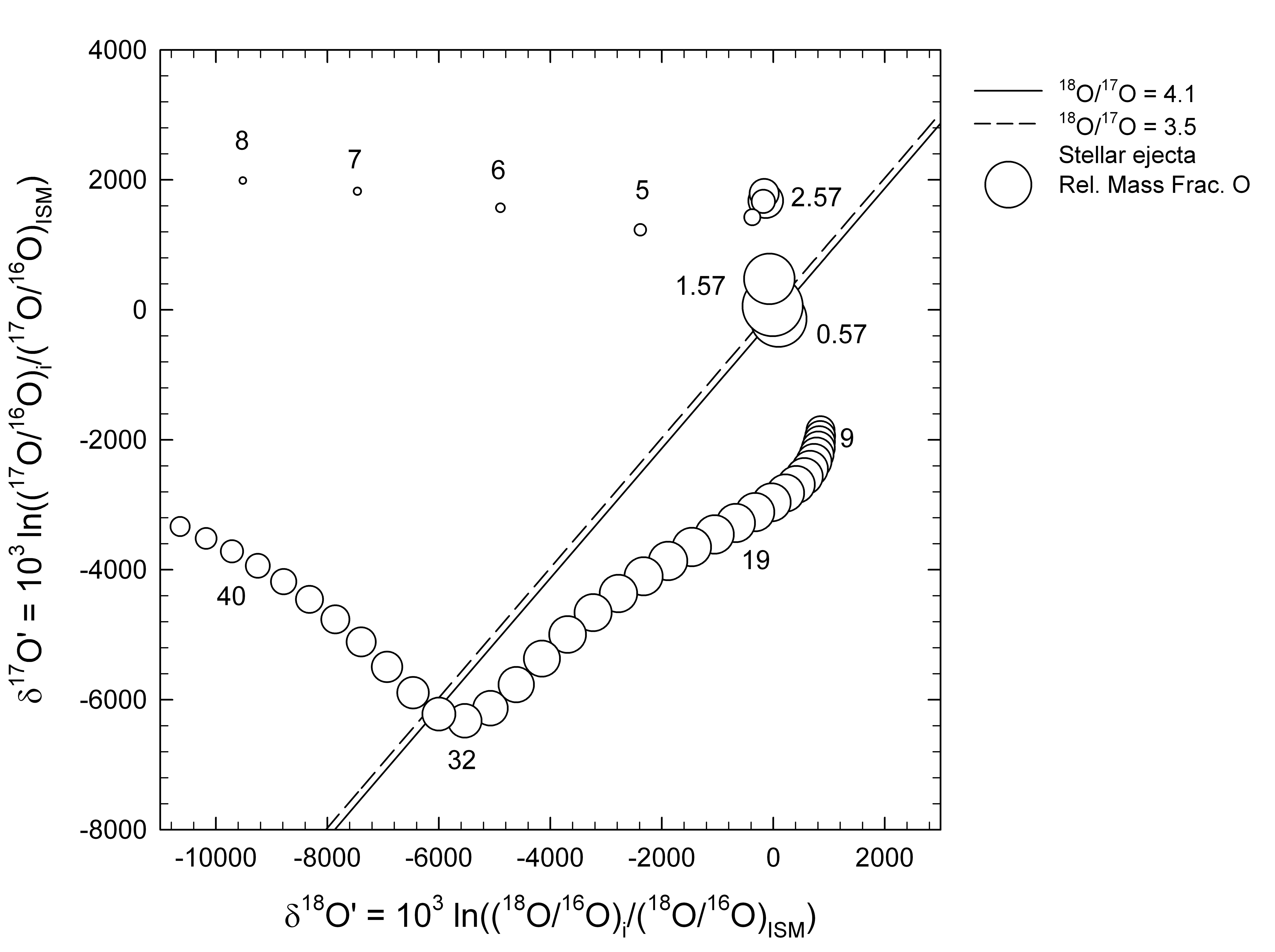}
\caption{Three-isotope plot showing the isotopic compositions of oxygen ejected by type II supernovae ($M_* > 8M_{\odot}$) and AGB stars ($M_* \le 8M_{\odot}$) obtained by smoothing supernova yields from RHHW02 with $\rm [^{17}O]$ production scaled to match that of WH07.  Numbers adjacent circles indicate progentior masses ($M_{\odot}$).  Each circle represents a progenitor mass bin of 1$M_{\odot}$.  The diameters of the circles represent the fractions of the total oxygen released by all stars comprising the complete IMF that are attributable to the individual mass bins (Equation (7)).  For this calculation a ``generic'' IMF was used where $\alpha_1 = 1.25$ for $M_*< 1M_{\odot}$, $\alpha_2 = 2.1$ for $1M_{\odot} \le M_*\le 2M_{\odot}$,  $\alpha_3 = 2.3$ for $2M_{\odot} < M_*\le 10M_{\odot}$, and $\alpha_4 = 2.3$ for $10M_{\odot} < M_*$.  The result is not altered significantly by employing a different IMF.  } 
\label{SNe_Yields}
\end{figure}

Using the calculations described above we can plot the isotopic compositions and mass fractions of oxygen ejected from stars comprising a stellar cluster in three-isotope space (Figure \ref{SNe_Yields}).  The result shows that SNe II produce a wide range of compositions from the high $\rm [^{18}O]/[^{16}O]$- low $\rm [^{17}O]/[^{16}O]$ ejected by smaller progenitors with masses less than 30$M_{\odot}$ to the low $\rm [^{18}O]/[^{16}O]$- high $\rm [^{17}O]/[^{16}O]$ ejected by the more massive progenitors.  It is clear that exogenous oxygen from low-mass SNe II ($<20M_{\odot }$) could explain the anomalously high $\rm [^{18}O]/[^{17}O]$ of the Sun compared with the more normal compositions of the Galaxy.  The implication is that the one or more stars that exploded and enriched the protosolar cloud with $\rm [^{18}O]/[^{16}O]$-rich oxygen were B stars.  The more massive O stars, on the other hand, eject oxygen with $\rm [^{18}O]/[^{16}O]$ too low to allow for mixing with ancient ISM to produce the solar oxygen isotope ratios while AGB stars, the sources of pre-solar grains in meteorites, produce oxygen too low in $\rm [^{18}O]/[^{17}O]$ to explain the solar values.  

\begin{figure}
\centering
\includegraphics[scale=0.35]{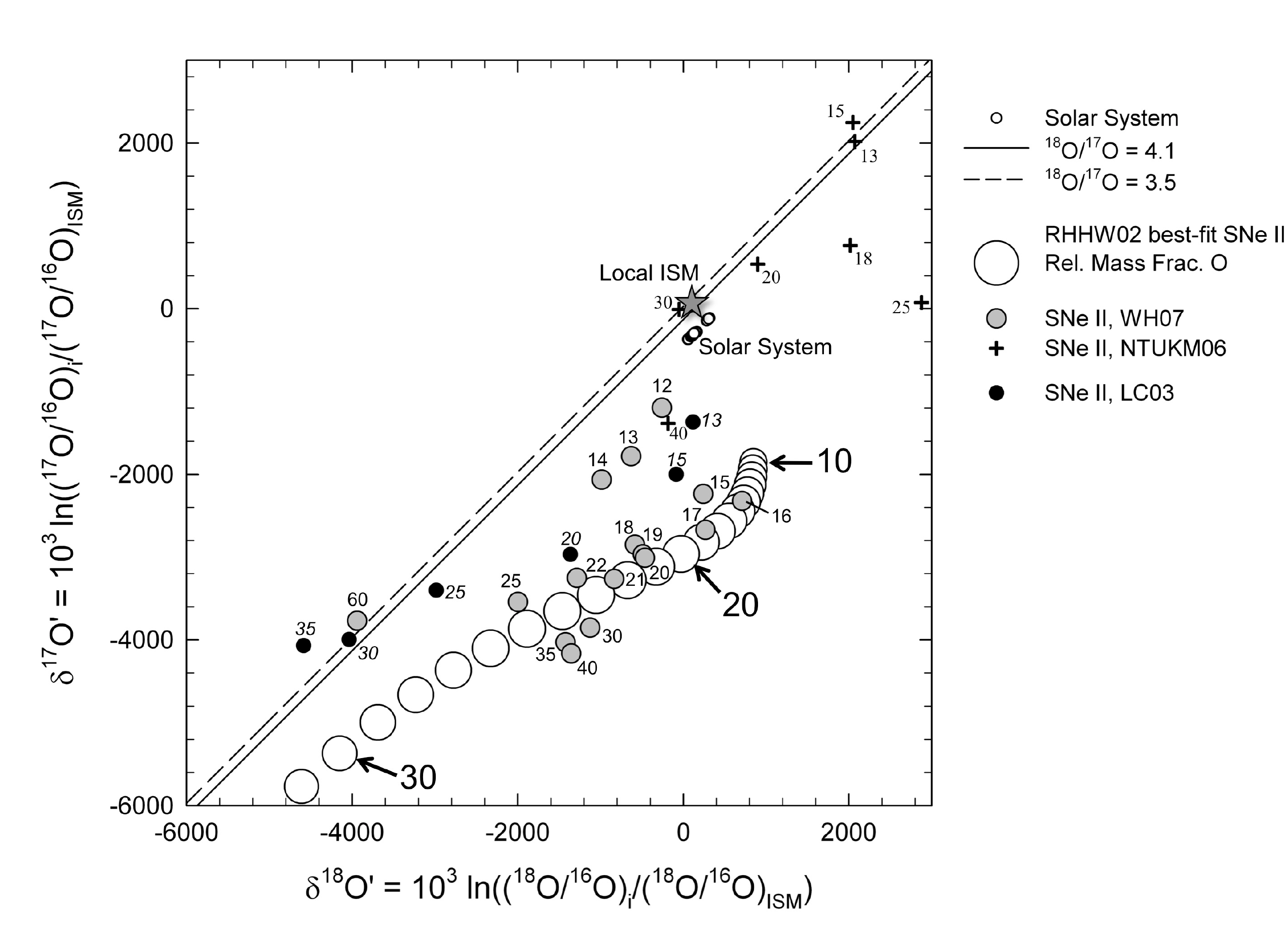}
\caption{Oxygen isotope ratios of SNe II ejecta as functions of progenitor mass ($M_{\odot}$) as predicted by LC03, NTUKM06, and WH07 for initial compositions of solar metallicity.    The fit used in the present study (Figure \ref{SNe_Yields}) is shown for comparison.  Also shown are data representing the solar system and the position of the local ISM as adopted in this paper.  Numbers next to each datum are the progenitor masses ($M_{\odot}$).  } 
\label{SN_yields_comparison}
\end{figure}

Supernova yield calculations are inherently uncertain and this uncertainty is difficult to quantify.  In order to illustrate the level of uncertainty, we show in Figure \ref{SN_yields_comparison} examples of ejecta oxygen isotope ratios as  functions of SNe II progenitor masses from a variety of recent studies.  In all cases one sees that the lower-mass progenitors in all of the calculations do indeed tend to produce higher $\rm [^{18}O]/[^{16}O]$ and $\rm [^{17}O]/[^{16}O]$ ejecta than the higher mass-progenitors.  However the exact locations of the yields relative to local ISM in three-isotope space vary considerably from model to model.  Of the calculations considered here, those of \cite{Nomoto2006-777} (NTUKM06) exhibit the greatest disparity relative to the others, with substantially greater abundances of both secondary oxygen nuclides compared with the results of WW95, RHHW02, \cite{Limongi2003-592} (LC03), and WH07.  Our smoothed yield vs. mass function is consistent with the most recent of these models (WH07) by design (Figure \ref{SN_yields_comparison}).  The veracity of what follows depends on the extent to which current calculations faithfully represent supernova oxygen isotope yields, but in all cases the inverse relationship between $\rm [^{18}O]/[^{16}O]$ and SN II progenitor mass is robust. 

We note that there is a discrepancy in the production of $\rm ^{18}O$ between WH07 and RHHW02.  The yield given by WH07 is substantially greater than that given by RHHW02 for stellar masses greater than $25$ $M_{\odot}$ (Figure \ref{SN_yields_comparison}).  However, we have performed the calculations described below using fits to both sets of oxygen isotope yields and find no appreciable difference in the results.  This is in part because we are concerned with supernova progenitor masses considerably less than $25$ $M_{\odot}$.  

For the purposes of these calculations, the precise form of the initial mass function (IMF) is unimportant.  This is because the main discrepancies between different IMFs are for the very low masses and the very high masses, while, as we show below, we are primarily concerned with the mass range from 8 to 30$M_{\odot}$. 

Given the apparent requirement that exogenous oxygen that affected the isotopic composition of the solar system came from B stars but not O stars,  the question then arises as to how oxygen from exploding B stars would be  ``selected'' by the solar precursor in greater abundance than oxygen from other sources.  The answer lies in the stochastic nature of star formation in general, as described below.  

\subsection{Oxygen Isotope Ratios of Ejecta as a Function of Cluster Size}
The relative age of the cluster proximal to the proto-solar molecular cloud is limited by the constraint on the average supernova progenitor star mass.  B stars having masses $<20M_{\odot } $ and $>$ 8$M_{\odot } $  (the minimum mass to produce a type II SN)  require 10  to 30 Myrs to evolve prior to explosion as SNe II \citep{Prantzos2008-311,Schaller1992-269}.  The stars that were the source of the high $\rm [^{18}O]/[^{17}O]$ oxygen therefore belonged to a generation of star formation that predated the solar system by at least 10 Myr.  We emphasize this conclusion.  The oxygen isotope data indicate an episode of enrichment that involves a generation of star formation that predated that which produced the Sun.  This conclusion is in contrast to, and inconsistent with, earlier models for supernova enrichment  of the solar system \citep[e.g., ][]{Schramm1982-236} in which it is assumed that the exploding supernova and the Sun formed together as siblings in the same star cluster.   

Enrichment of a region of star formation by explosions of B stars from an earlier generation of star formation is consistent with protracted star formation in molecular cloud complexes in the Galaxy spanning 10 to 20 Myrs \citep[e.g.,][]{Gounelle2009-1}.  With this scenario of propagating star formation in mind, we used a statistical analysis to examine the likelihood for $\rm [^{18}O]/[^{17}O]$ enrichment of the proto-solar molecular cloud by oxygen ejected from B stars that evolve to become SNe II .  We adopted 20 Myr as a conservative upper limit for the time interval over which B stars reside in a cloud complex after formation, corresponding to a minimum stellar mass of 11 $M_{\odot}$.  Current evidence indicates that young clusters of moderate size that produce one or more B stars disrupt their parental molecular clouds on time scales of $\sim 3$ to 10 Myr \citep{Leisawitz1989-731,Elmegreen2007-1064}, a time span less than the 10 Myr required for the most massive, and therefore the most short-lived, B stars (18$M_{\odot } $, $\sim$ B0)  to explode as type II supernovae.  However, molecular clouds are usually present in cloud complexes extending over hundreds of parsecs, and these complexes are sites of protracted episodes of star formation lasting tens of millions of years \citep{Elmegreen2007-1064,Hartmann2001-852}.  With velocity dispersions of $\sim 10$ km/s between stellar subgroups within a complex \citep{Bruijne1999-585}, the more massive B stars (B0 to B1) can drift only 100 to 200 pc during their lifetimes.  Such a star cannot leave the vicinity of the giant molecular cloud (GMC) complex before exploding as a SN II; B0 to B1 stars formed in one cloud are likely to be in close proximity to other clouds within the same extended complex when they become supernovae.  Indeed, such supernovae are often invoked as triggers for successive generations of star formation \citep{Briceno2007-345}. Evidence that star clusters can encounter multiple clouds is provided by the presence of multiple generations of stars within some well studied clusters \citep{Mackey2009-275}.  The net result is that complexes hundreds of pc in size persist for up to 50 Myr, GMCs within the complexes survive for 10 to 20 Myr, and GMC cores that produce individual clusters last $\sim 3$ Myr \citep{Elmegreen2007-1064}. 

\begin{figure}
\centering
\includegraphics[scale=0.06]{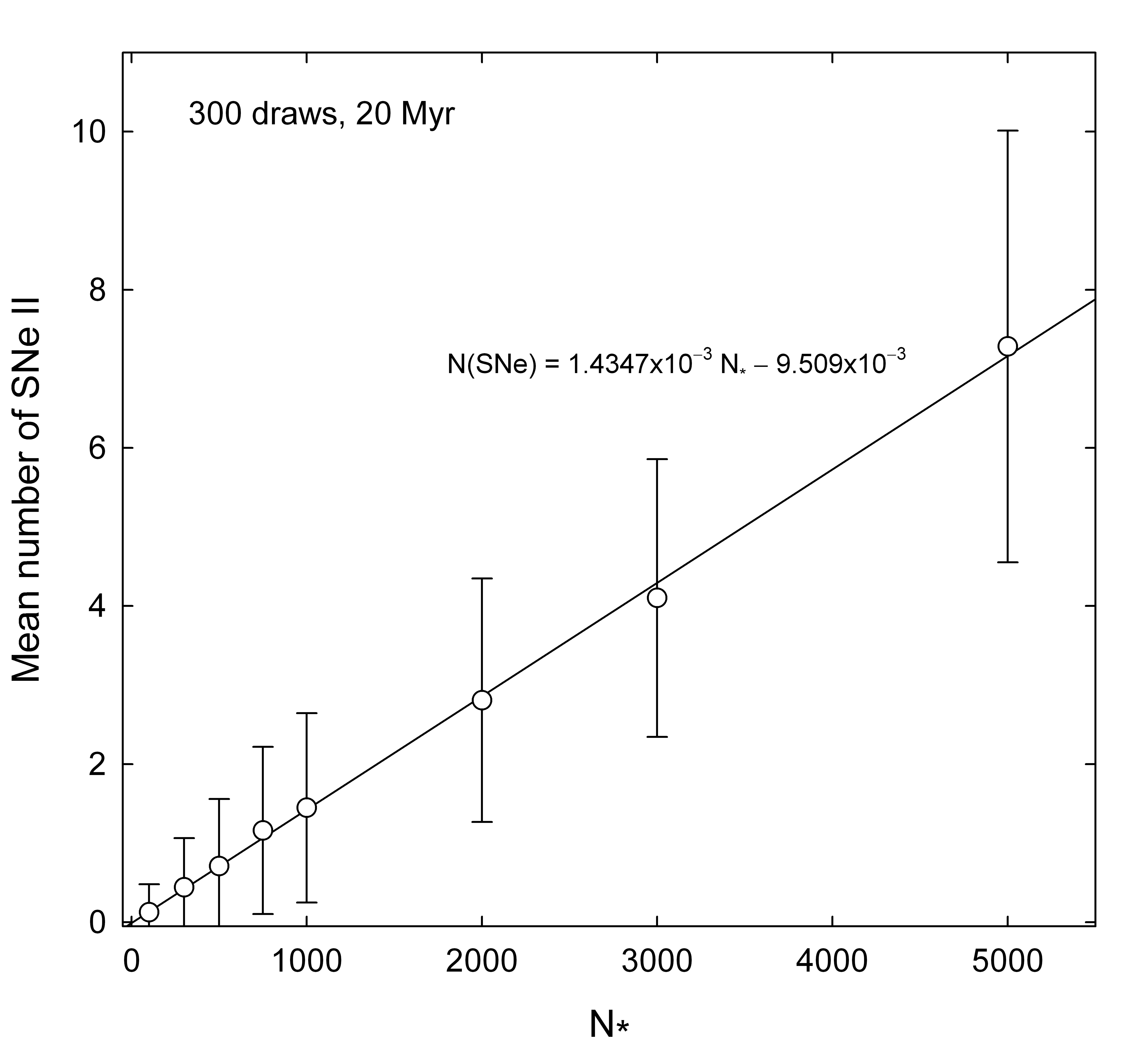}
\caption{Plot of mean number of type II SNe (ordinate) occurring within 20 Myr of birth of star cluster versus number of stars comprising the cluster (abscissa) produced by 300 random draws for each mass interval using equation (\ref{mass_generation}).  The 20 Myr constraint corresponds to a minimum stellar mass of 11 $M_{\odot}$.  Error bars represent the 1$\sigma$ distribution for the random draws at each cluster size.  All cluster sizes produce the same fractional number of stars that become SNe II ($1.4 \times 10^{-3}$), showing that all of the clusters obey random sampling of the same initial mass function. } 
\label{Mean_SNe_vs_N}
\end{figure}

\begin{figure}
\centering
\includegraphics[scale=0.1]{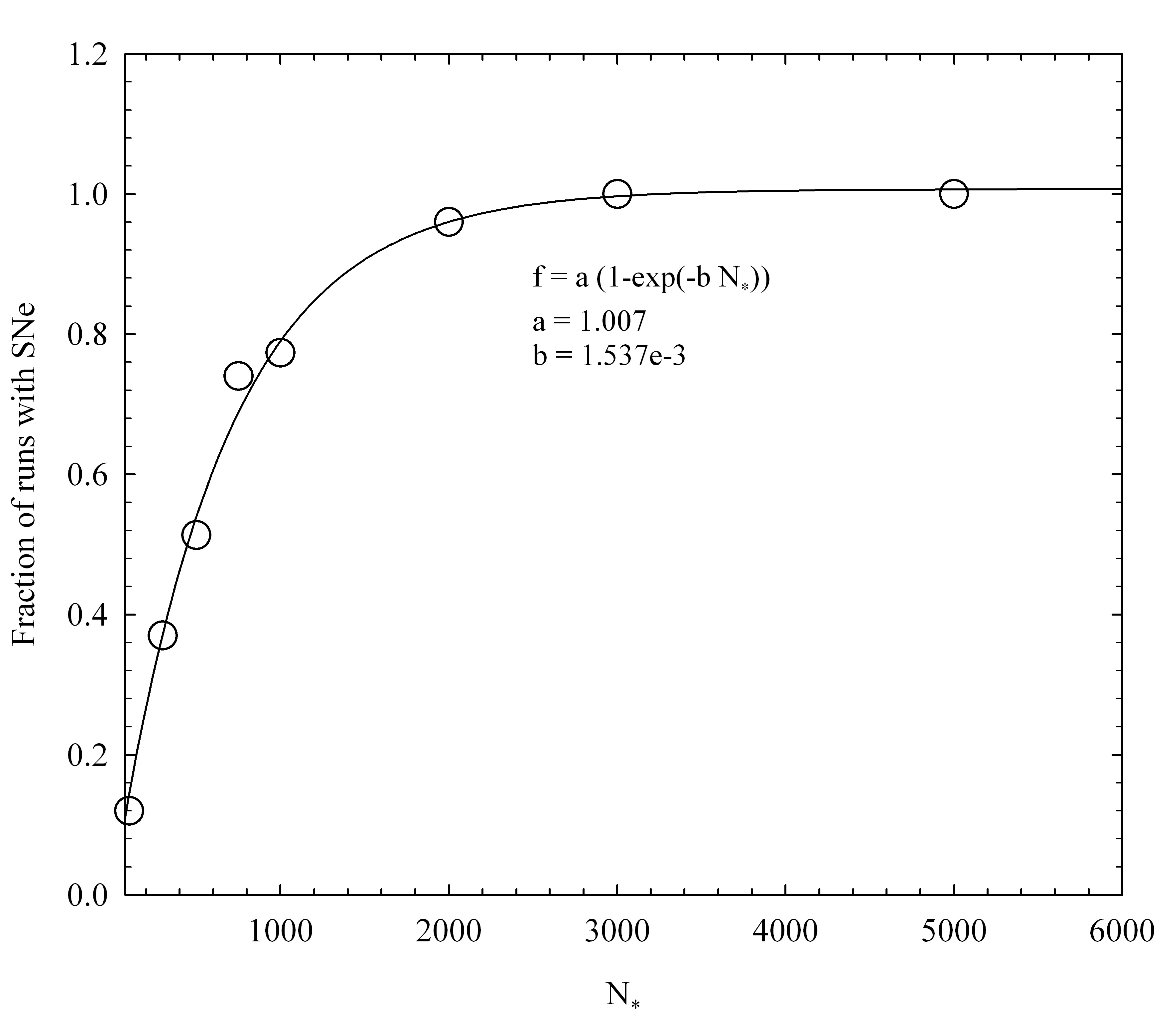}
\caption{Plot of fraction of clusters composed of $N_*$ stars that produce SNe II within 20 Myr of formation of the cluster as produced using equation (\ref{mass_generation}).  A fit to these simulation results provides an expression for the the fraction of clusters that produce supernovae II, $f$, as a function of number of stars comprising the cluster: $f=(1-exp(-1.54\times10^{-3} N_*))$. } 
\label{frac_SNe_20Myr_fit}
\end{figure}

\begin{figure}
\centering
\includegraphics[scale=0.4]{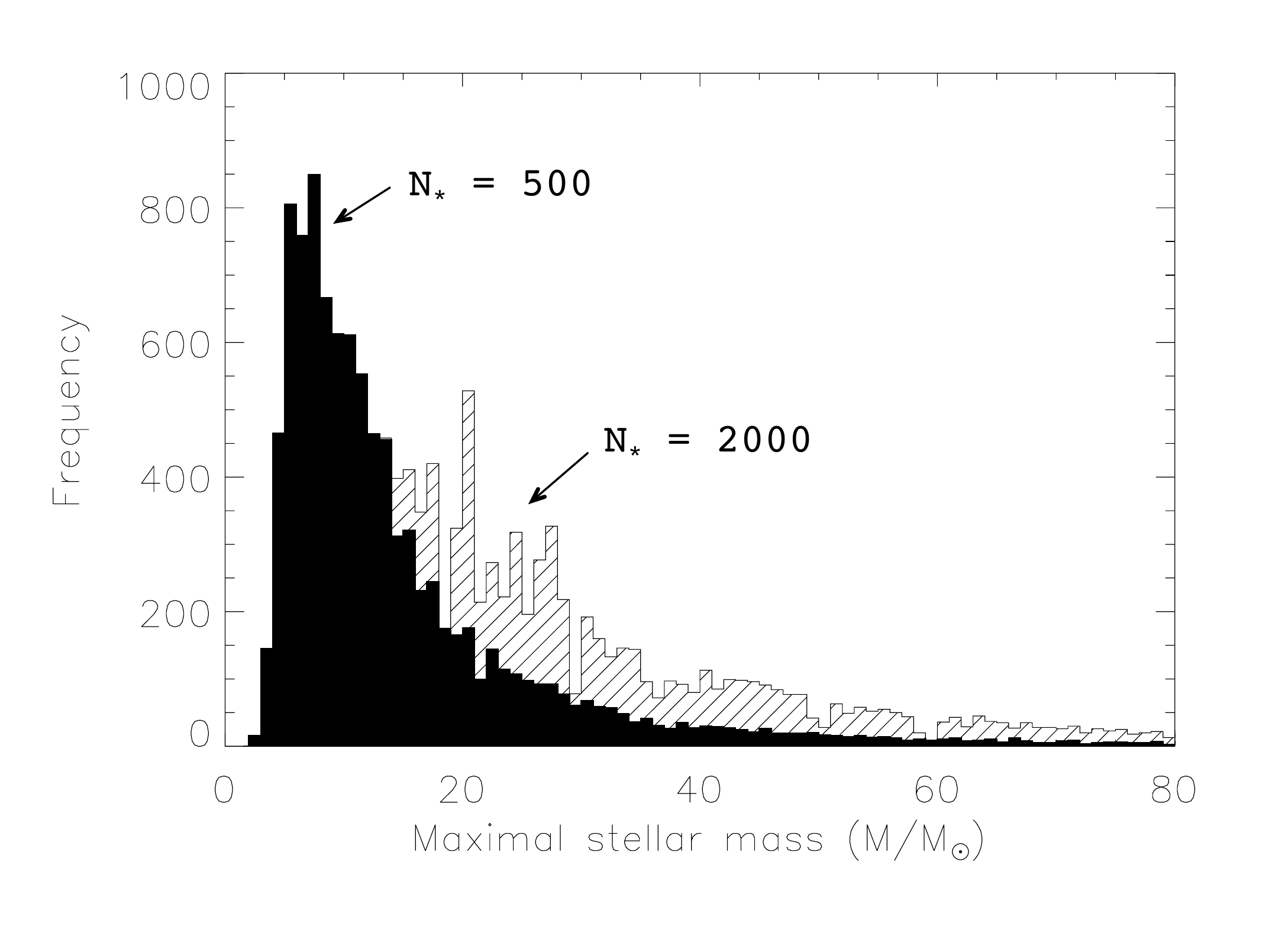}
\caption{Histogram showing the frequency of occurrences of the maximal stellar mass produced in each of $10^4$ random instances of clusters composed of 500 stars (solid) and 2000 stars (lined).  Note that the mass of the largest star tends to be greater for the larger clusters.  In particular, 
the peak in the mass of the largest star shifts from $~8$  $M_{\odot}$ for $N_* = 500$  to approximately 20 $M_{\odot}$ for $N_* = 2000$.    } 
\label{histogram_max_sizes}
\end{figure}

For the statistical analysis we used the mass generation function of \cite{Kroupa1993-545} modified by \cite{Brasser2006-59} to obtain 300 random realizations of star clusters of various sizes ranging from tens to tens of thousands of members.   With this function the mass of the \textit{j}th star of the cluster is obtained from the expression

\begin{equation} 
M_{j}/M_{\odot} =0.01+(0.19x^{1.55} +0.05x^{0.6} )/(1-x)^{0.58}  
\label{mass_generation}
\end{equation} 

\noindent where \textit{x} is a uniformly distributed random number between 0 and 1.  Sampling of the IMF, as simulated  with equation (\ref{mass_generation}), results in systematic relationships between occurrences of SNe II, the maximum size of SNe progenitors, and cluster size.  Figure \ref{Mean_SNe_vs_N} shows that the average fractional number of stars that produce type II SNe within 20 Myr of the birth of the cluster is $1.4\times 10^{-3}$ for all cluster sizes.  The fact that all of the simulated clusters produce the same average fractional number of supernovae illustrates that clusters of all sizes do indeed represent random samplings of the IMF.  However, the discrete, stochastic sampling produces important differences in populations as a function of cluster size.  Clusters composed of relatively few numbers of stars can sometimes produce no SNe II at all simply because the progenitor stars (those with $M_*$ $>$  $8 M_{\odot}$) are relatively rare.  This is illustrated in Figure \ref{frac_SNe_20Myr_fit} where it can be seen that, with the 20 Myr time constraint where we consider only stars with masses greater than 11 $M_{\odot}$, only about 15 \% of clusters composed of 100 stars, a practical minimum size for clusters \citep{Lada2003-57},  produce SNe II whereas 80\% of clusters of 1,000 stars produce SNe II in these simulations.  The maximum stellar mass, and therefore most massive supernova progenitor, also varies with cluster size as a fundamental consequence of discrete sampling of the IMF in which the more massive the star the more rare its occurrence.  As cluster size increases, the maximum in the frequency of most massive members of the cluster shifts to higher mass (Figure \ref{histogram_max_sizes}).  In other words, smaller clusters tend to produce smaller supernovae and larger clusters tend to produce larger supernovae.  

We emphasize two important points in this regard.  First, the histograms in Figure \ref{histogram_max_sizes} do not depict the frequency distributions of all masses, only that of the most massive member of each cluster.  Second, that the propensity for larger clusters to more reliably produce the more massive stars is a natural consequence of stochastic sampling of any initial mass function in which frequency varies inversely with stellar mass.   Although debate surrounds whether sampling of the IMF by star clusters is truly random \citep{Weidner2010-275} , the positive correlation between the mode in maximal star mass and cluster size is evident nonetheless \citep[e.g., ][figure 4]{Weidner2010-275}.   

We combined supernova oxygen isotope yields with the statistical analysis described above in order to examine the relationship between star cluster size and the oxygen isotopic composition of the oxygen ejected by SNe II from the cluster.  The results are depicted in oxygen three-isotope space by contouring the relative probability of occurrence of supernova oxygen ejecta of a given isotopic composition.  The isotope ratios of the ejecta are integrated from time 0 to 20 Myr after the (instantaneous) formation of the cluster.   Assignment of probabilities is an expediency that amounts to a Gaussian smearing of each model datum.  In this way clusters of adjacent points receive greater weight (greater probability) than individual points.  For this purpose each of 300 time-integrated oxygen isotopic compositions of SNe II ejecta is smeared by a Gaussian distribution:

\begin{equation}
P_{{}^{17}{\text{O}}}  = \frac{1}
{{\sigma \sqrt {2\pi } }}\exp \left( { - \frac{1}
{2}\left( {\frac{{\delta {}^{17}{\text{O}'} - \delta {}^{17}{\text{O}'}_{{\text{Model}}} }}
{\sigma }} \right)^2 } \right)
\end{equation}
 
and

\begin{equation}
P_{{}^{18}{\text{O}}}  = \frac{1}
{{\sigma \sqrt {2\pi } }}\exp \left( { - \frac{1}
{2}\left( {\frac{{\delta {}^{18}{\text{O}'} - \delta {}^{18}{\text{O}'}_{{\text{Model}}} }}
{\sigma }} \right)^2 } \right)
\end{equation}

\noindent where ${\delta {}^{17}{\text{O}'}_{{\text{Model}}} }$ and ${\delta {}^{18}{\text{O}'}_{{\text{Model}}} }$ are the model supernova ejecta isotope ratios in $\delta'$ notation and $\sigma$ is taken to be slightly larger than the grid spacing.  The normalized joint probability for a given grid square in three-isotope space is then

 \begin{equation}
dP_{{\text{grid}}}  = \frac{1}{n}{P_{{}^{{\text{17}}}{\text{O}}} P_{{}^{{\text{18}}}{\text{O}}} \;d\delta {}^{17}{\text{O}'}\;d\delta {}^{18}{\text{O}'}}
\end{equation}
 
 \noindent where $d\delta^{17}{\text{O}'} = d\delta^{18}{\text{O}'} = \sigma$ and $n$ is the number of points (300 in this case).  For the results shown here we set $\sigma$ equal to a convenient but arbitrary level of smearing of 40\permil  \, compared with a grid spacing of 33\permil.  Altering $\sigma$ and the grid spacing has no substantive effect on the results.  
 
\begin{figure}
\centering
\includegraphics[scale=0.07]{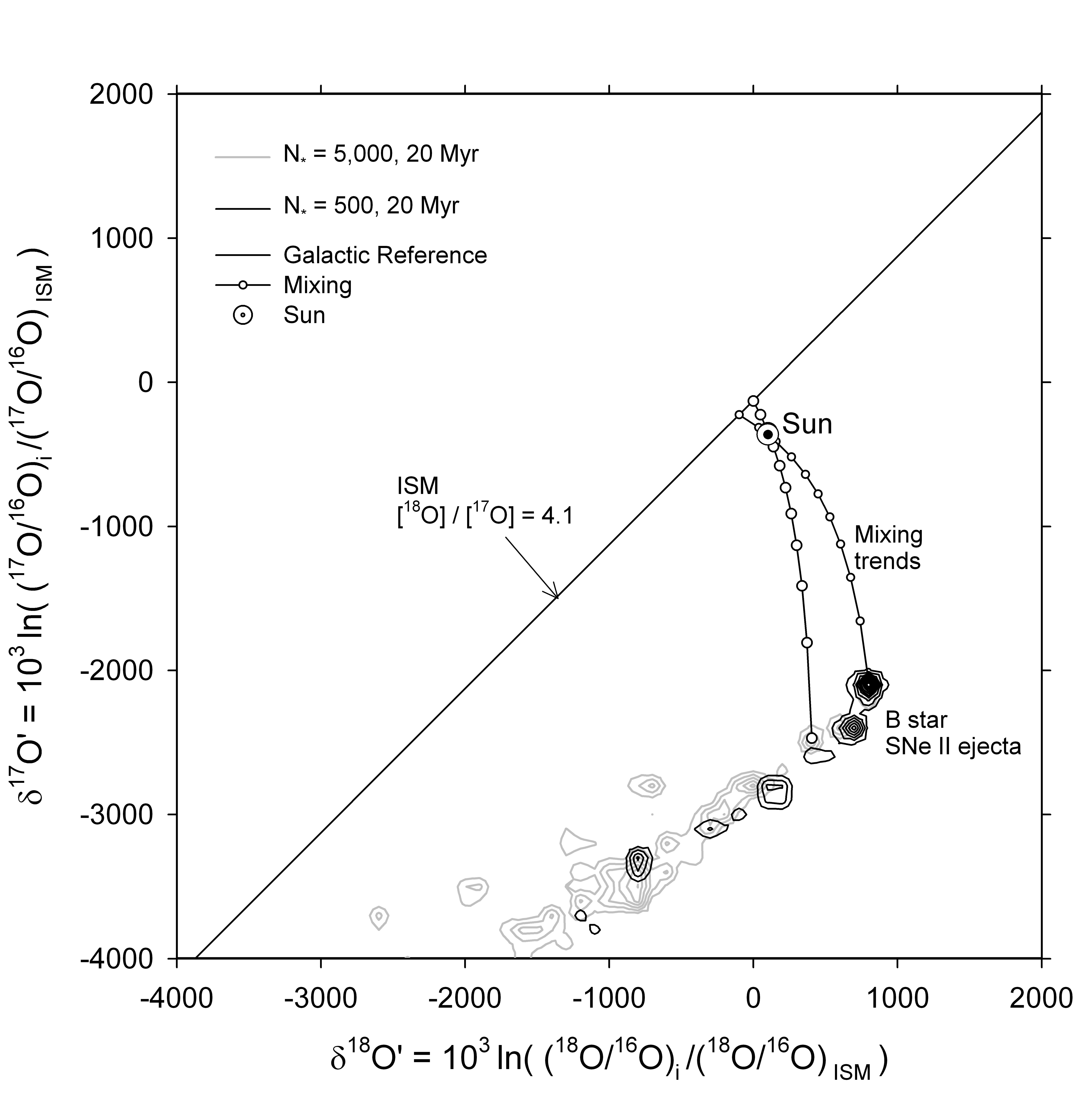}
\caption{The ISM ($\rm [^{18}O]/[^{17}O]$ = 4.1), Sun (circle/dot), and example results from the statistical analysis (contours) shown in oxygen three-isotope space. Black contours show the  probability density for 300 realizations of the oxygen isotopic compositions of ejecta from clusters of 500 stars emitted over a 20 Myr time interval. Grey contours show the probability density for ejecta from clusters of 5,000 stars.  The contour interval is 0.005 to a maximum of 0.04.  The two mixing curves show the range of possible ways of mixing SNe II ejecta with ISM oxygen 4.6 Gyr before present having a typical Galactic $\rm [^{18}O]/[^{17}O]$ of 4.1 to form the Sun and solar system with $\rm [^{18}O]/[^{17}O]$ of 5.2.}
\label{Prob_contours_Bstars}
\end{figure}

Probability density contours for the isotopic compositions of oxygen ejected over a 20 Myr period from clusters of 500 and 5,000 stars are shown in Figure \ref{Prob_contours_Bstars}.  The contours quantify the inverse relationship between cluster size and $\rm [^{18}O]/[^{16}O]$ of  oxygen ejected by SNe II.  Oxygen ejected {\it en masse} from a cluster of  5,000 stars (grey contours) is more likely to extend to lower $\rm [^{18}O]/[^{16}O]$ than is oxygen ejected from a cluster composed of 500 stars (black contours).  

We also show in Figure \ref{Prob_contours_Bstars} calculated mixing curves between estimates for the ISM 4.6 Gyr before present and the most likely compositions produced by a cluster of ca. 500 stars that would produce the composition of the Sun.  One curve is based on mixing with the most probable composition for a 500 star cluster.  This peak in the probability density for the 500 star cluster occurs at the highest $\rm [^{18}O]/[^{16}O]$ permitted by the 20 Myr constraint (i.e., the lowest possible mass for SNe II progenitors).   The other mixing curve is based on the lowest $\rm [^{18}O]/[^{16}O]$ ratio consistent with the oxygen isotopic composition of the ISM 4.6 Gyr before present (i.e., the present-day $\rm [^{18}O]/[^{16}O]$ ISM, see above).  The mixing curves demonstrate that smaller clusters, represented here by our statistical representation of clusters of 500 stars, were more likely sources of oxygen isotope enrichment for the solar system than larger clusters (e.g.,  the 5,000 star clusters represented by grey contours).  

\begin{figure}
\centering
\includegraphics[scale=0.075]{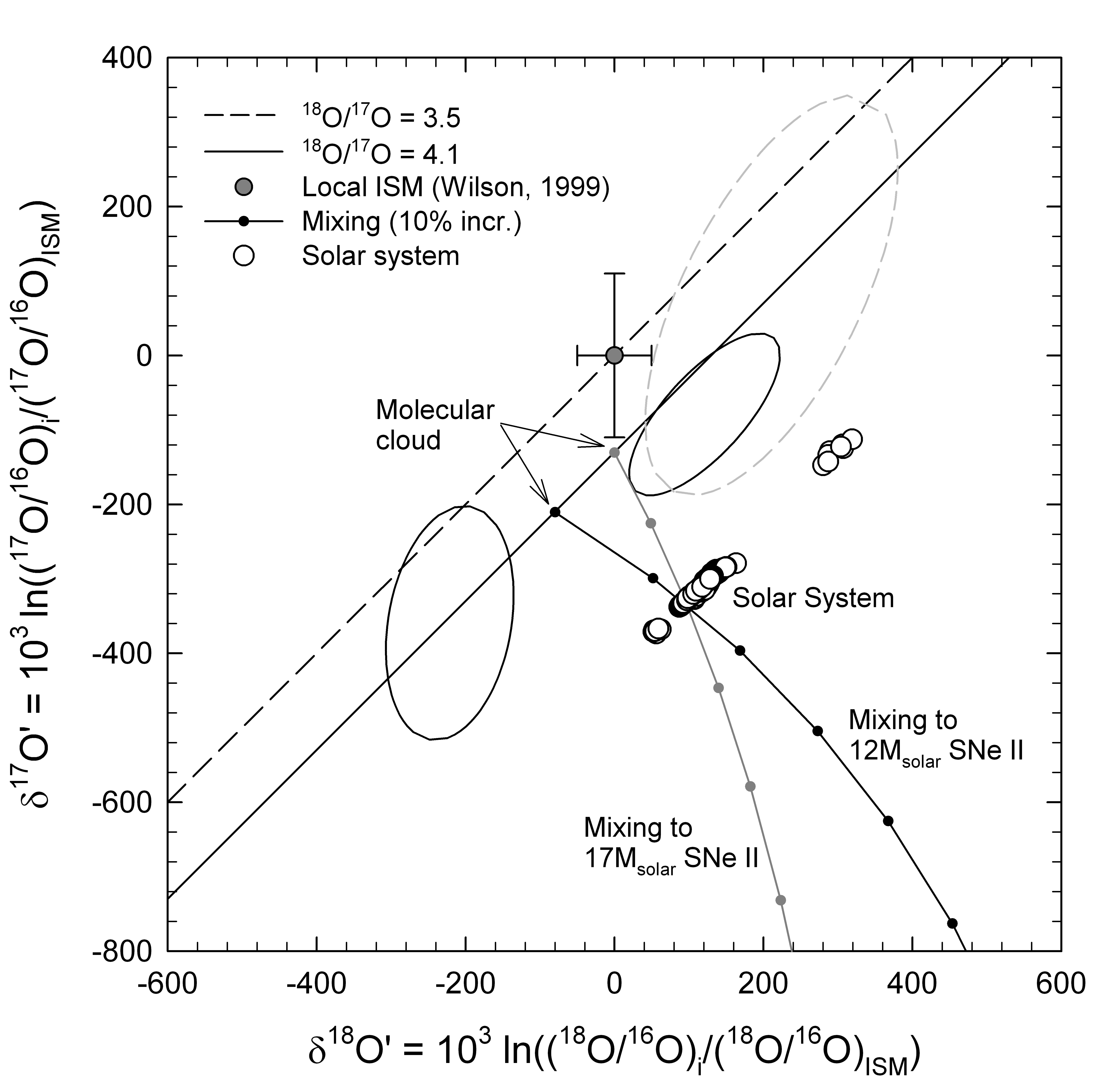}
\caption{Three-isotope ratio plot showing the 95\% confidence ellipses for YSOs VV CrA and RE 50 and the YSO IRAS 19110+1045 datum.  Errors in the IRAS 19110+1045 data are dashed because they are poorly characterized in $\rm [^{18}O]/[^{16}O]$  vs. $\rm [^{17}O]/[^{16}O]$ space as a consequence of using a curve-of-growth analysis of limited precision to obtain the $\rm C^{16}O$ column density.   Dots on the ISM-SNe II mixing curves represent 10\% increments of addition of SNe II oxygen to molecular cloud oxygen with  $\rm [^{18}O]/[^{17}O] = 4.1$.  The two curves are the same as those shown in Figure \ref{Prob_contours_Bstars}.  The steeper mixing curve corresponds to mixing with ejecta from an average SNe II progenitor mass of 17$M_{\odot } $ while the shallower curve shows mixing with debris from a SNe II progenitor mass of 12 $M_{\odot } $.}\, 
\label{data_3isotope}
\end{figure}

Results of this analysis show that oxygen ejecta with the composition required to explain the oxygen isotope ratios of the solar system relative to the ISM 4.6 Gyr before present is more than twice as likely to have come from a star cluster of several hundred stars than from a cluster of several thousand stars.  Larger clusters tend to produce oxygen too low in $\rm [^{18}O]/[^{16}O]$ while considerably smaller clusters produce too few SNe II.  In our analysis, 19\% of clusters of 500 stars produce oxygen isotope ratios suitable to explain the solar composition by supernova enrichment.  For comparison, the corresponding probability for a cluster of 5,000 stars is 7\%.  
Relationships between the mxing curves constrained by the contours in Figure \ref{Prob_contours_Bstars}, the oxygen isotopic composition of the ISM, and the oxygen isotopic composition of the solar system are shown in Figure \ref{data_3isotope}. 

The mass fraction of exogenous oxygen in the solar parental cloud can be estimated from the mixing curves in Figures \ref{Prob_contours_Bstars} and \ref{data_3isotope}.  The masses of ejected oxygen represented by the two mixing curves in Figures \ref{Prob_contours_Bstars} and \ref{data_3isotope} are 0.7 and 7.5 $M_{\odot } $ for the high and low $\rm [^{18}O]/[^{16}O]$ cases, respectively.  The minimum mass of the enriched proto-solar cloud material, $M_{\rm MC}$, can be calculated from the assumption of a 100\% efficiency of injection and simultaneous solution of the mass balance equations 

\begin{equation}
C_{{}^{17}{\text{O}}, \odot }  = \frac{{\left( {M_{{}^{17}{\text{O,MC}}}  + M_{{}^{17}{\text{O,EJ}}} } \right)}}
{{M_{{\text{MC}}} }}
\end{equation}

\begin{equation}
C_{{}^{18}{\text{O}}, \odot }  = \frac{{\left( {M_{{}^{18}{\text{O,MC}}}  + M_{{}^{18}{\text{O,EJ}}} } \right)}}
{{M_{{\text{MC}}} }}
\end{equation}

\noindent where $C_{^{17}{\rm O},\odot}$ and $C_{^{18}{\rm O},\odot}$ are the solar concentrations by mass of the specified isotopes, $M_{{}^{17}{\text{O,MC}}}$ ($M_{{}^{18}{\text{O,MC}}}$) is the unknown initial mass of $\rm ^{17}O$ ($\rm ^{18}O$) in the cloud material , and $M_{^{17}{\text{O,EJ}}}$ ($M_{^{18}{\text{O,EJ}} }$)  is the calculated mass of $\rm ^{17}O$ ($\rm ^{18}O$) added to the cloud by the SNe II ejecta constrained by the mixing curves.  With the additional constraint that the initial $\rm [^{18}O]/[^{17}O]$ of the molecular cloud was on the Galactic line, such that $M_{{}^{17}{\text{O,MC}}} =  M_{{}^{18}{\text{O,MC}}} (17/18)/4.1$, we obtain  $M_{\rm MC} \sim 700$ to 5,000 $M_{\odot }$.  The mass fraction of SNe II oxygen of 0.1\% (e.g., $100\times 0.7/700$), together with SNe II oxygen mass fractions 5 to 10 times solar \citep{Rauscher2002-323},  suggests a total contribution of SNe II ejecta to the proto-solar molecular cloud  of $\sim 1$\% by mass.  

\subsection{Collateral Isotopic Effects}
Enrichment of the proto-solar cloud by SNe II ejecta may help to explain other puzzling aspects of the isotopic composition of the solar system. We consider here the implications for the isotopes of silicon, carbon, and the short-lived radionuclides $\rm ^{60}Fe$ and $\rm ^{26}Al$ as well as for the interpretation of pre-solar oxide grains.  A caveat to any such calculation that attempts to use a single supernova source for numerous isotopic and elemental systems is the prospect for heterogeneous mixing between the supernova ejecta and the target cloud material.  Nonetheless, 
in order to assess the collateral implications of $~1\%$ enrichment of the proto-solar molecular cloud by SNe II ejecta for other isotope systems, we derive a general mixing equation that illustrates explicitly the relative importance of elemental abundances and isotope ratios in producing mixtures of molecular cloud and SNe ejecta.  Mass balance for the number of atoms of isotope 1 of element $E$, $n_{E_1}$, between pre-enrichment molecular cloud (MC) material, supernova ejecta (SNe) and the final mixture (solar, $\odot$) can be written in terms of the total atoms for each reservoir $j$, $N_j$, and the atomic fractions of nuclide $E_1$ in reservoir $j$, $(x_{E_1})_j$, such that

\begin{eqnarray}
(x_{E_1 } )_ \odot   &=& \frac{{n_{E_1 } }}{{N_{{\text{MC}}}  + N_{{\text{SNe}}} }} \nonumber \\
&=& (x_{E_1 } )_{{\text{MC}}} x_{{\text{MC}}} + (x_{E_1 } )_{{\text{SNe}}} x_{{\text{SNe}}} 
\label{xE1}
\end{eqnarray}

\noindent where $x_{{\text{SNe}}}  = N_{{\text{SNe}}} /(N_{{\text{SNe}}}  + N_{{\text{MC}}} )$, the atomic fraction of supernova ejecta in the mixture, and $x_{\rm MC}$ is the atomic fraction of original molecular cloud material.  Recognizing that $x_{\rm MC} = 1-x_{\rm SNe}$ and taking the ratio of equation (\ref{xE1}) for two isotopes of element $E$ we obtain

\begin{eqnarray}
{}^{2/1}R_ \odot   &=& \frac{{(x_{E_2 } )_ \odot  }} {{(x_{E_1 } )_ \odot  }} \nonumber \\
&=& \frac{{x_{{\text{SNe}}} (x_{E_2 } )_{{\text{SNe}}}  + (1 - x_{{\text{SNe}}} )(x_{E_2 } )_{{\text{MC}}} }}
{{x_{{\text{SNe}}} (x_{E_1 } )_{{\text{SNe}}}  + (1 - x_{{\text{SNe}}} )(x_{E_1 } )_{{\text{MC}}} }}
\label{Rdot}.
\end{eqnarray}

\noindent  Equation (\ref{Rdot}) is rearranged to obtain an expression for the atomic fraction of supernova ejecta in terms of elemental abundances and isotope ratios for supernova ejecta, pre-enrichment molecular cloud, and the final solar mixture:

\begin{equation}
x_{{\text{SNe}}}  = \frac{{(x_{E_1 } )_{{\text{MC}}} \left( {{}^{2/1}R_ \odot   - {}^{2/1}R_{{\text{MC}}} } \right)}}
{\Gamma }
\label{XSNe}
\end{equation}

\noindent where

\begin{eqnarray}
\Gamma  &=& (x_{E_1 } )_{{\text{SNe}}} {}\, ^{2/1}R_{{\text{SNe}}}  - (x_{E_1 } )_{{\text{MC}}} {}\, ^{2/1}R_{{\text{MC}}}  \nonumber \\
&-& {}^{2/1}R_ \odot  \left( {(x_{E_1 } )_{{\text{SNe}}}  - (x_{E_1 } )_{{\text{MC}}} } \right).
\label{denom}
\end{eqnarray}

\noindent  In practice we can equate atomic fractions with atoms per hydrogen in applying equation (\ref{XSNe}).  We will assume that the relative abundances of the elements (as opposed to isotope ratios)  in the precursor cloud were indistinguishable from solar values.  Numerical experiments in which fictive molecular cloud elemental abundances were used confirm that deviations from this simplifying assumption have negligible effects on the results. 

\subsubsection{Carbon isotope ratios}
The solar system $\rm [^{12}C]/[^{13}C]$ of 89 is greater than the typical present-day local ISM value of $\sim 68$ \citep{Milam2005-1126}.  However, the majority of this difference is explained by the increase in $\rm ^{13}C$ relative to $\rm ^{12}C$ by GCE over the last 4.6 Gyr \citep{Prantzos1996-760}.  Application of equation (\ref{XSNe}) using the s19 SNe II model of RHHW02 ($\rm [^{12}C]/[^{13}C]  = 193.9$) shows that addition of $1\%$ by mass of SNe ejecta from B stars would have raised the pre-enrichment molecular cloud $\rm [^{12}C]/[^{13}C]$ from $~85$ to the solar value of 89.  There appear to be no observational constraints against such a shift. 
 
 \subsubsection{Silicon isotope ratios}
\begin{figure}
\centering
\includegraphics[scale=0.07]{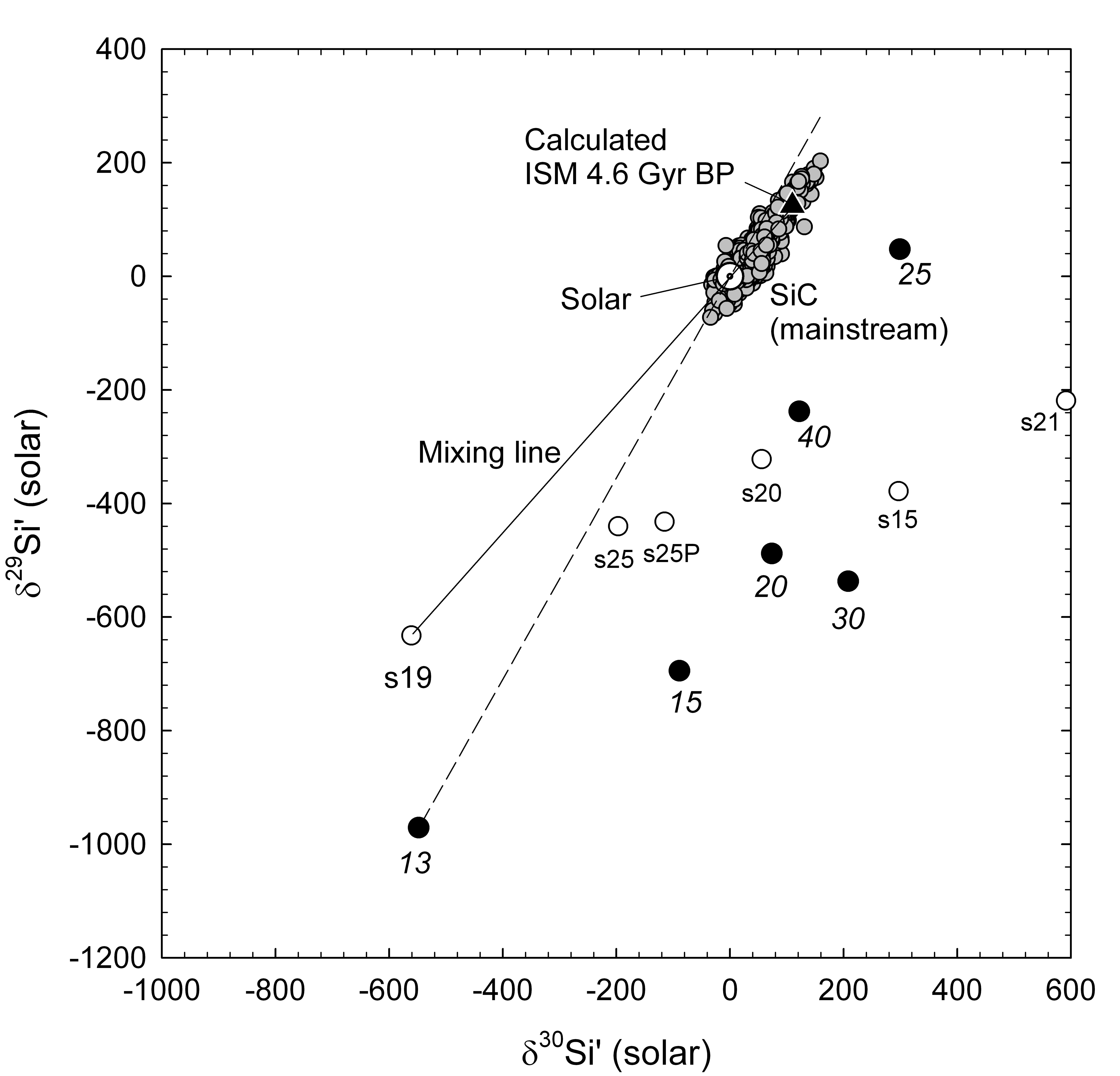}
\caption{Silicon three-isotope ratio plot in which $\rm [^{29}Si]/[^{28}Si]$ (ordinate) and $\rm [^{30}Si]/[^{28}Si]$ (abscissa) are normalized to solar ratios of $0.05078$ and $0.03347$, respectively.  The plot shows the relative positions of the Sun ($\odot$), the pre-solar mainstream SiC grains (grey circles, Ernst Zinner, pers. comm.), the various supernovae ejecta models of RHHW02 (open circles), and supernovae ejecta models of NTUKM06 (solid circles).  SNe models are labeled with progenitor masses ($M_{\odot}$).  Results of a mixing calculation yielding the composition of the interstellar medium 4.6 Gyr before present (black triangle) based on mixing $1\%$ by mass of ejecta from the s19 model of RHHW02 with the ISM to produce the solar composition is shown with the solid line.  The dashed line shows the trajectory of the mixing that would be produced using the 13$M_{\odot}$  progenitor SNe ejecta of NTUKM06.  }
\label{Silicon}
\end{figure}

A long-standing problem has been an apparent excess in $\rm ^{28}Si$ in the solar system relative to expected values.  An excess in $\rm ^{28}Si$ over $\rm ^{29}Si$ and $\rm ^{30}Si$ in the solar system relative to the ISM 4.6 Gyr before present is evident by comparisons with pre-solar SiC grains \citep{Alexander1999-222}.  Mainstream SiC grains come from AGB stars that predate the Sun by hundreds of millions to billions of years.  Galactic chemical evolution should therefore have resulted in the younger Sun having greater $\rm [^{29}Si]/[^{28}Si]$ and $\rm [^{30}Si]/[^{28}Si]$ than these earlier-formed AGB stars, yet solar values are lower, not higher, by $~11$ to $12\%$ \citep{Alexander1999-222}.  Type II supernovae expel an excess of $\rm ^{28}Si$ relative to the heavier Si isotopes \citep{Rauscher2002-323} and it has been suggested previously that the solar system might have been enriched by $\rm ^{28}Si$ from supernovae \citep{Alexander1999-222}.  Models for SNe II ejecta tend to produce too much $\rm ^{30}Si$ relative to $\rm ^{29}Si$ compared with the silicon isotopic compositions necessary to explain the relationship between solar system and mainstream SiC pre-solar grains (Figure \ref{Silicon}), but this may be a problem of spurious overproduction of $\rm ^{30}Si$ in the calculations \citep{Alexander1999-222}.   Using equation (\ref{XSNe}), we find that addition of $1\%$ by mass of ejecta from the s19 model of RHHW02  decreases $\rm [^{29}Si]/[^{28}Si]$ and $\rm [^{30}Si]/[^{28}Si]$ in the pre-enrichment solar system molecular cloud by $12\%$ and $11\%$, respectively, making the ISM 4.6 Gyr before present greater in $\rm [^{29}Si]/[^{28}Si]$ and $\rm [^{30}Si]/[^{28}Si]$ than the majority of pre-solar mainstream SiC grains.  Therefore, the same enrichment process that explains the aberrant  $\rm [^{18}O]/[^{17}O]$ of the solar system could also explain most of the  excess in $\rm ^{28}Si$ in the solar system (Figure \ref{Silicon}).  A caveat is that the oxygen isotopic composition of the s19 supernova ejecta model is not ideal for explaining the anomalous $\rm [^{18}O]/[^{17}O]$ of the solar system as it is slightly lower in $\rm [^{18}O]/[^{16}O]$ than progenitors depicted with the mixing cuves in Figures \ref{Prob_contours_Bstars} and \ref{data_3isotope}, and would therefore require  the ISM 4.6 Gyr before present to have been higher in $\rm [^{18}O]/[^{16}O]$ than today (violating expectations from GCE).  This problem is not severe, however, given the uncertain dispersion in ISM $\rm [^{18}O]/[^{16}O]$ along the Galactic slope-1 line in three-isotope space. 

\subsubsection{Short-lived radionuclides}
Assessing the effects of SNe II enrichment on the abundances of short-lived radionuclides is complicated by the opportunity for decay of these radioisotopes in the interval between SNe ejection and formation of the first solids in the solar system (the so-called ``free decay" time).  In these cases SNe II ejecta isotope ratios ${^{i/j}\!R}$ are modified to account for decay of radionuclide $i$ such that ${^{i/j}\!R} = {^{i/j}\!R}_{\rm SNe II}\,  \exp(-\Delta t/ \tau)$ where $\tau$ is the mean life of nuclide $i$ and $\Delta t$ is the free decay time.   \cite{Gounelle2009-1} have argued that a process of enrichment from an earlier generation of star formation can explain the short-lived radioactive nuclides that are evinced in the solar system.  Our oxygen isotope results place constraints on possible progenitors of these nuclides.  For example, addition of $1\%$ by mass of SNe II debris from one or more B0 to B1 stars, as suggested by the $\rm [^{18}O]/[^{17}O]$ data and represented by model s19 of RHHW02 ($\rm [^{60}Fe]/[^{56}Fe] = 8.3x10^{-4}$, $\rm [^{56}Fe]/[H] = 2.61x10^{-4}$), would yield the amount of $\rm ^{60}Fe$ (mean life = $2.2$ Myr) in the solar system ($\rm [^{60}Fe]/[^{56}Fe] \sim 5x10^{-7}$, $\rm [^{56}Fe]/[H] = 3.16x10^{-5}$) inferred from meteorites \citep{Gounelle2008-781} with a free decay time between nucleosynthesis and incorporation to solar system precursor cloud material of $\sim 11$ Myr. By this time shorter-lived $\rm ^{26}Al$ and $\rm ^{41}Ca$, two nuclides that appear to correlate in the solar system \citep{Sahijpal1998-559}, would have decayed away, requiring that the origin of these species was separate from that of the anomalous abundances of the stable isotopes and $\rm ^{60}Fe$.  

\subsubsection{Presolar grains}
\begin{figure}
\centering
\includegraphics[scale=0.07]{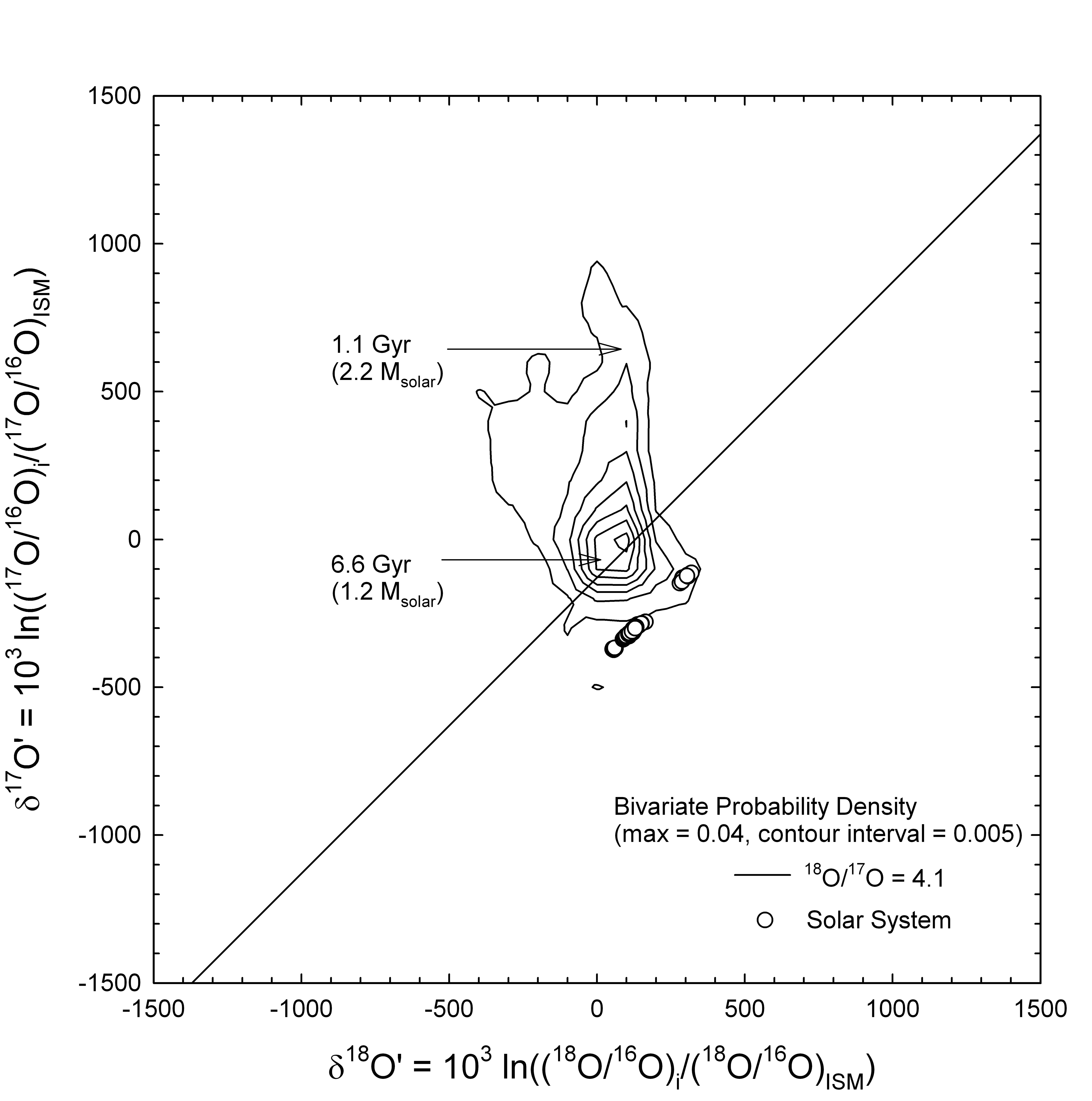}
\caption{Probability contour plot of the oxygen isotopic compositions of pre-solar oxygen-bearing grains (data compilation from Larry Nittler, pers. comm.).  The peak in the density of pre-solar oxygen isotope ratios coincides with the typical Galactic $\rm [^{18}O]/[^{17}O]$ value of 4.1, represented by the line in the figure.  This $\rm [^{18}O]/[^{17}O]$ is  significantly lower than the solar value (solar system is shown as open circles).  Models \citep{Nittler2009-271} for the masses and ages of the AGB sources of these grains suggest that they span many billions of years of Galactic evolution. }\, \, 
\label{presolar_grains}
\end{figure}

The oxygen isotope ratios of pre-solar oxide grains found in meteorites bear on the question of the GCE of oxygen.  \cite{Nittler2009-271} argues that models for the origins of these grains  are consistent with  $\rm [^{18}O]/[^{17}O]$ equal to the solar value of 5.2 for all of the AGB stars from which they derive .  In all cases, however, fundamental problems arise with interpretations of pre-solar grain oxygen isotope ratios when the solar system $\rm [^{18}O]/[^{17}O]$ is not equal to the interstellar medium (ISM) today  \citep[e.g., ][]{Alexander1999-222}.  Nittler and colleagues have interpreted the pre-solar oxides as having come from a range of masses of AGB stars from about 1.2 to 2.2 $M_{\odot}$.  This corresponds to a sampling of Galactic oxygen over  ~5.5 Gyrs based on the lifetimes of the progenitors stars between 6.6 and 1.1 Gyr \citep[e.g.,][]{Schaller1992-269}.  That is to say, a 1.2 $M_{\odot}$ star that enters the AGB phase of evolution at the time the solar system was forming sampled oxygen from the ISM 6.6 Gyr prior to the formation of the Sun, or by any measure of stellar lifetimes, very early in the evolution of the Galaxy.  A 2.2 $M_{\odot}$ star entering the AGB phase at the time the Sun was forming represents a sampling of oxygen from ~1.1 Gyr prior to the formation of the Sun.  Therefore, if the interpretation of the pre-solar grain data is taken at face value, the ISM was characterized by a solar $\rm [^{18}O]/[^{17}O]$ of 5.2 for ~7 billion years leading up to the formation of the Sun, then sometime between 4.6 Gyr ago and now, the bulk $\rm [^{18}O]/[^{17}O]$ of the ISM changed by nearly 30\% after having been constant for the prior 7 Gyrs.  Such a change might be expected once the Milky Way ages  to the point where high-mass stars are no longer being made, but that is not the case yet. Therefore, we conclude that it may be necessary to modify models for the origins of pre-solar grains.  We note that the  pre-solar grain oxygen isotope data cluster about a Galactic $\rm [^{18}O]/[^{17}O]$ of about 4.1 (Figure \ref{presolar_grains}), suggestive of a causal relationship between the peak in the pre-solar grain data and the Galactic $\rm [^{18}O]/[^{17}O]$ (though a coincidence cannot be discounted).  In addition,  there are pre-solar oxide grains that fall below the solar $\rm [^{18}O]/[^{17}O]$ line (with higher than solar $\rm [^{18}O]/[^{17}O]$) that are not consistent with AGB or supernova predictions and so are as yet unexplained by any model, suggesting that the origin of high $\rm [^{18}O]/[^{17}O]$  grains may not be understood in general. 

\subsection{Probability of Clusters with Suitable Oxygen Isotope Ratios }
The peculiar nature of the solar $\rm [^{18}O]/[^{17}O]$ suggests that the type of chemical and isotopic enrichment during star formation proposed here is not the norm.  The likelihood for the enrichment by B stars but not O stars can be estimated from the probability for the occurrence of clusters of stars having the requisite number of stars  (ca. 500 as opposed to several thousand) and the fraction of those clusters that produce supernovae.  The  cluster mass function in terms of the distribution ($f$) of cluster sizes (specified by the number of stars comprising the cluster,  $N_*$, rather than mass for consistency with our statistical analysis) is characterized with a power law \citep[e.g.][]{Elmegreen1997-480, Parmentier2008-678} such that

\begin{equation}
\frac{{df(N_* )}}
{{dN_* }} \propto N_*^{ - \alpha } .
\label{freq}
\end{equation}

\noindent The distribution function is therefore  $f(N_*) \propto N_*^{1-\alpha}$.  The likelihood of a cluster having  from $N_{\rm min}$ to $ N_{\rm max}$ stars is given by the integral of the distribution function:

\begin{equation}
P_{\Delta N_* }  = a\int_{N_{\min } }^{N_{\max } } N_*^{1-\alpha }{dN_* } 
\label{prob}
\end{equation}

\noindent  where $a$ is the normalizing factor. The value for $\alpha$ in equation (\ref{freq}) that characterizes the cluster mass function is generally found to be near $2.0$ \citep{Elmegreen1997-480,Haas2010-512}.  In this case $a=0.109$ for a total range in $N_*$ from 100 to $10^6$.  However, reported values for $\alpha$ vary from about 1.8 to 2.4 \citep{Haas2010-512} with $a$ varying accordingly.  For clusters of 100 to 1,000 members and $\alpha=2.0$, $P_{\Delta N_{*} } = 0.25$.  For $\alpha$ ranging from $1.8$ to $2.4$ the range in $P_{\Delta N_{*} } $ is $0.11$ to $0.62$.

The mass generation function used here (equation (\ref{mass_generation})) yields a best-fit relationship between the number of stars comprising a cluster and the fraction of those clusters that produce supernovae within the 20 Myr timescale (Figure \ref{frac_SNe_20Myr_fit}):
\begin{equation}
X_{\rm SNe} =1-\exp (-bN_{*} )
\label{supernovae}
\end{equation} 

\noindent where $b = 1.537\times10^{-3}$. From equations (\ref{prob}) and (\ref{supernovae}), the joint probability for occurrences of SNe II within 20 Myr from a cluster of an appropriate range in $N_*$ is
\begin{equation}
P_{\Delta N_* , {\rm SNe}}=\int_{N_{\min } }^{N_{\max } } {a N_*^{ 1-\alpha} \left( 1-\exp ( - bN_* )\right)} \;dN_* 
\label{joint}
\end{equation}

\noindent Evaluation of equation (\ref{joint}) yields $P_{\Delta N_* , {\rm SNe}} = 0.10$ for $\alpha = 2.0$ with a range due to uncertainty in $\alpha$ from 0.05 ($\alpha = 1.8$) to 0.23 ($\alpha=2.4$).  Our 5\% to 20\% probability is consistent with previous estimates for the fraction of stars formed under the influence of massive stars adjacent molecular clouds \citep{Mizuno2007-128,Hennebelle2007}.  We must also consider, however, that not every cluster in this size range that produces supernovae has the right oxygen isotopic composition because ocassionally some of these clusters do produce the larger O stars that are too low in $\rm [^{18}O]/[^{16}O]$ to be consistent with enrichment of the proto-solar cloud.  The simulation shown in Figure \ref{Prob_contours_Bstars}  suggests that  roughly half of the clusters of ca. 500 stars that produce supernovae also have B stars as their maximal stellar mass and so produce oxygen isotope ejecta suitable to explain the solar system $\rm [^{18}O]/[^{17}O]$.   The probability derived above should therefore be halved, yielding an estimate of $~2.5$ to $10\%$ for the occurrence of a suitable star cluster.  These conclusions are invalid if SNe II oxygen isotope yields are closer to the high-$\rm [^{18}O]/[^{16}O]$ values calculated by NTUKM06 rather than the values reported by RHHW02, LC03 and WH07, for example.

\subsection{The Potential for Enrichment by WR Winds }
Another source of oxygen to consider in a star-forming region is that produced by the prodigious winds of Wolf-Rayet stars (in particular WC stars with He burning products exposed at the surface).  We conclude that this is a less likely alternative to SNe II ejecta.  Firstly, WR stars evolve from more massive O stars generally \citep{Crowther2007-45}, making them by their very nature less common than less massive B stars that end their lives as type II supernovae.  Indeed, the ratio of the rates of occurence of type II SNe to the rates of occurrence of type I b/c supernovae, the likely endpoint for WR stars, is about 5  \citep{Vanbeveren2005-0286}.  Secondly, the WR phase of evolution lasts for $10^5$ years \cite[e.g.][ and references therein]{Maeder1994-287} and so the chances of catching winds from this phase of evolution alone, without also capturing the ensuing collapse supernova debris, are small.  Thirdly, rates of mass loss from WR stars are of order $10^{-4}$ to $10^{-5}$ $M_{\odot}$ /yr, and these rates multiplied by the duration of the WR phase of evolution yields of order one solar mass of total wind material \citep[e.g.][]{Binns2006-47}.  The mass of oxygen released will be considerably less than the total mass in these winds.  Therefore, the mass of oxygen released is $<<1$ $M_{\odot}$  compared with the minimum of approximately 1 $M_{\odot}$ of oxygen liberated by low-mass type II supernovae.  When considering the brevity of the WR phase of evolution, the relative rarity of O star WR progenitors, and the relatively low oxygen yields from the winds, it seems that enrichment of a star forming region in oxygen isotopes  from WC stars is not as likely as enrichment by SNe II ejecta. 

\section{CONCLUSIONS}
We conclude that the comparison between solar and extra-solar oxygen isotope ratios in YSOs provides a signature of enrichment of the proto-solar molecular cloud by supernovae from an earlier generation of star formation.   Proximity to a previous generation of star formation implies that the Sun formed in a molecular cloud complex where multiple generations of star formation occurred.  The source of exogenous oxygen that enriched the solar system parental cloud is more than twice as likely to have come from a smaller cluster of several hundred stars than from a larger cluster of several thousand stars.  Tighter constraints on the oxygen isotopic compositions of ejecta from type II SNe would considerably improve our ability to use isotope ratios to infer the birth environment of the solar system. 
 
\acknowledgments
This work is based on observations collected at the European Southern Observatory Very Large Telescope under program ID 179.C-0151.  The authors acknowledge sponsorship from the NASA's Origins Program (EDY, MRM), the NASA Astrobiology Institute (EDY, MRM),  the Programme National de Plan{\'e}tologie (PNP), the CNRS fund France-{\'E}tats-Unis, and the European program ORIGINS [grant
MRTN-CT-2006-035519] .  KMP was provided support by NASA through Hubble Fellowship grant no. 1201.01 awarded by the Space Telescope Institute. The authors are indebted to  Geoff Blake (Caltech) for access to the NIRSPEC data for IRAS 19110+1045 and to Bruce Elmegreen (IBM) for enlightening communications.

\appendix
\section{Solar system $\rm [^{18}O]/[^{17}O]$}

Specifying a single $\rm [^{18}O]/[^{17}O]$ ratio for the solar system has three sources of uncertainty.  One is systematic errors associated with measuring absolute ratios of isotopes.  A second is the dispersion in this ratio within the solar system.  Yet another is the difficulty in obtaining oxygen isotope ratios for the Sun itself.  We address these sources of uncertainty here.  

{\it Absolute ratio:}
The $\rm [^{18}O]/[^{16}O]$ of standard mean ocean water (SMOW), the commonly accepted reference for oxygen isotope ratios, was determined by  \cite{Baertschi1976-31} to be $0.00205.2 \pm 0.0000045$ (1$\sigma$), corresponding to a $\rm [^{16}O]/[^{18}O]$ of $498.7 \pm 0.1$.  This value has remained undisputed and was determined mass spectrometrically using mixtures of $\rm D_2^{18}O$ and $\rm H_2^{16}O$.  The $\rm [^{17}O]/[^{16}O]$  for SMOW is less precisely known.  Values over the years have generally ranged from $0.000365$ to $0.00040$ \citep{Santrock1985-1444}, corresponding to $\rm [^{16}O]/[^{17}O]$  values of 2739 to 2500.  The range in $\rm [^{18}O]/[^{17}O]$ of SMOW based on the well-defined $\rm [^{16}O]/[^{18}O]$ of SMOW and the full range in reported  SMOW $\rm [^{16}O]/[^{17}O]$ values is from 5.0 to 5.5.  Recently, the $\rm [^{17}O]/[^{16}O]$ of SMOW was determined relative to an absolute $\rm [^{13}C]/[^{12}C]$ for the so-called Pee Dee Belemnite standard using measurements of $\rm CO_2$ isotopologue ratios \citep{Assonov2003-1017} .  The Pee Dee Belemnite (PDB) is a marine fossil composed of calcite that serves as the primary reference for terrestrial carbon isotope studies.  The PDB $\rm [^{13}C]/[^{12}C]$ used for this study is  $0.011237 \pm 0.000030$ as determined by \cite{Craig1957-133}, corresponding to $\rm [^{12}C]/[^{13}C]$ = $89.0 \pm 0.2$.  The $\rm [^{17}O]/[^{16}O]$ SMOW value determined by \cite{Assonov2003-1017}  tied to the \cite{Craig1957-133} PDB carbon isotope ratio is $0.0003867 \pm 0.0000009$, or  a $\rm [^{16}O]/[^{17}O]$ of $2586 \pm 6$.  Using this latest value for the absolute abundance of $\rm ^{17}O$, the  $\rm [^{18}O]/[^{17}O]$ of SMOW is $5.18 \pm 0.01$ ($1\sigma$).  

However, the PDB carbon isotope ratio has an associated uncertainty that affects the determination of $\rm [^{18}O]/[^{17}O]$ for SMOW  \citep{Assonov2003-1007,Assonov2003-1017}.   An alternative to relying on a reference carbon isotope ratio is to determine $\rm [^{17}O]/[^{16}O]$  and $\rm [^{13}C]/[^{12}C]$ simultaneously for a well-characterized sample of $\rm CO_2$.   \cite{Valkiers2007-10}  performed this experiment on $\rm CO_2$ gas generated by acid digestion of a commonly used carbonate reference material, NBS-19 calcite.  For this $\rm CO_2$ gas these authors obtained $\rm [^{17}O]/[^{16}O] = 0.00038014 \pm (48)$, $\rm [^{18}O]/[^{16}O] = 0.00208365 \pm (48)$, and $\rm [^{13}C]/[^{12}C] = 0.0111593 \pm (16)$.  Because the difference in carbon isotope ratios between NBS-19 and PDB is precisely known, the NBS-19 $\rm CO_2$ carbon isotope ratio corresponds to a $\rm [^{13}C]/[^{12}C]$ for PDB of $0.0111376 \pm (16)$, or $\rm [^{13}C]/[^{12}C] = 89.79 \pm 0.02$ .  The differences in oxygen isotope ratios between this NBS-19 $\rm CO_2$ gas and SMOW are mass dependent.  It is therefore possible to back out the $\rm [^{17}O]/[^{16}O]$ of SMOW from these new data and the relationship

 \begin{equation}
 \frac{{\left( {\frac{{{}^{{\text{17}}}{\text{O}}}}
{{{}^{{\text{16}}}{\text{O}}}}} \right)_{{\text{NBS - 19,CO}}_{\text{2}} } }}
{{\left( {\frac{{{}^{17}{\text{O}}}}
{{{}^{{\text{16}}}{\text{O}}}}} \right)_{{\text{SMOW}}} }} = \left( {\frac{{\left( {\frac{{{}^{{\text{18}}}{\text{O}}}}
{{{}^{{\text{16}}}{\text{O}}}}} \right)_{{\text{NBS - 19,CO}}_{\text{2}} } }}
{{\left( {\frac{{{}^{18}{\text{O}}}}
{{{}^{{\text{16}}}{\text{O}}}}} \right)_{{\text{SMOW}}} }}} \right)^\beta
 \label{massfrac}  
\end{equation}
 
 \noindent where the exponent $\beta$  characterizes the mass-dependent fractionation law.  Values for $\beta$ that relate mass-dependent  variations in $\rm [^{17}O]/[^{16}O]$ to those of $\rm [^{18}O]/[^{16}O]$ range from $0.51$ to $0.53$ \citep{Young2002-1095}.  For any value for $\beta$ in this range, equation \ref{massfrac}, the data from  \cite{Valkiers2007-10}, and the accepted $\rm [^{18}O]/[^{16}O]$ for SMOW yields a SMOW $\rm [^{17}O]/[^{16}O]$ value of $0.0003725$ ($\rm [^{16}O]/[^{17}O] = 2685$), well within the range of previous estimates.  The resulting $\rm [^{18}O]/[^{17}O]$ for SMOW is $5.38$.
 
One can conclude from the discussion above that the absolute $\rm [^{18}O]/[^{17}O]$ of SMOW, the primary oxygen reference for the solar system, is  $5.2$ based on the commonly accepted carbon isotope ratio for PDB, and that there is an uncertainty in the SMOW $\rm [^{18}O]/[^{17}O]$ of about $\pm 0.2$ as a result of the uncertainty in the absolute ratio of carbon isotopes in the PDB reference material.  We show below that the dispersion in the ratio of the rare isotopes of oxygen in the solar system is sufficiently small that the uncertainty in the absolute $\rm [^{13}C]/[^{12}C]$ for PDB is currently the dominant source of uncertainty in $\rm [^{18}O]/[^{17}O]$ for solar system rocks.   

\begin{figure}
\centering
\includegraphics[scale=0.7]{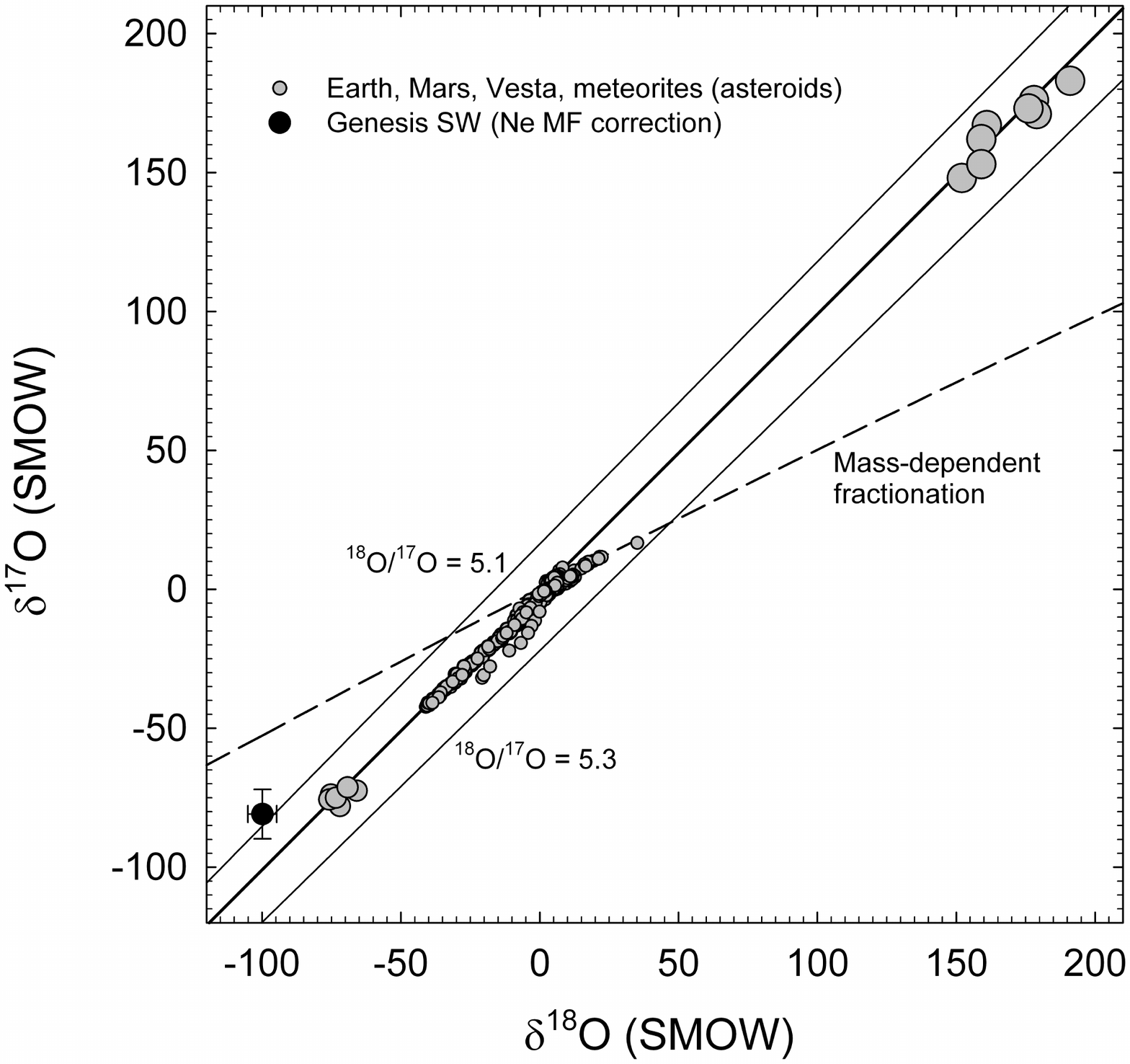}
\caption{Three-isotope plot relative to standard mean ocean water (SMOW) showing oxygen isotope ratios of meteorites and terrestrial rocks representing Earth, Mars, asteroid 4 Vesta, and the asteroid parent bodies represented by all classes of meteorites (grey circles).  The sources of this compilation of literature data include \cite{Clayton1977-209}, \cite{Clayton1984-48}, \cite{Clayton1991-2317}, \cite{Clayton1993-115}, \cite{Clayton1996-1999}, \cite{Young1998-452}, \cite{Young1999-1331}, \cite{Kobayashi2003-663}, and \cite{Sakamoto2007-231} among others.  With two exceptions, all data are results of fluorination of rocks, yielding the greatest precision.  The exceptions are the two extreme values obtained on rare materials by secondary ion mass spectrometry presented by \cite{Kobayashi2003-663} and \cite{Sakamoto2007-231}.   Uncertainties in each datum are smaller than the symbols used.  These data span the full range of compositions, including highly fractionated FUN inclusions as summarized by \cite{Krot2010-1159}.  The result for  solar wind from the Genesis mission \citep{McKeegan2010V31B-2316} with $2\sigma$ error bars is shown for comparison.   Also shown for comparison are lines of constant $\rm [^{18}O]/[^{17}O]$ based on a value for SMOW of $5.2$.}\, \, 
\label{appendixfig1}
\end{figure}

{\it Dispersion of oxygen isotope ratios in the solar system:}
Deviations in $\rm [^{18}O]/[^{17}O]$ from the value for SMOW arise as a result of mass-dependent fractionation of the oxygen isotopes.  Most physicochemical processes result in mass-dependent partitioning at some level.   At issue is how much fractionation exists in solar system materials.  Figure \ref{appendixfig1} shows the full range of oxygen isotope ratios for solar system rocks.  The most fractionated ratios with the highest $\rm \delta^{18}O$ are the result of low-temperature water-rock reactions and/or evaporation and are not representative of primordial solar system oxygen isotope ratios.  Nonetheless, it is clear that {\it all} of the data for solar system rocky bodies fall within $0.1$ of $\rm [^{18}O]/[^{17}O] = 5.2 $ regardless of level of fractionation.  This dispersion in $\rm [^{18}O]/[^{17}O]$ is less than the uncertainty imparted by ambiguity in absolute isotope ratios, as described above.

{\it Oxygen isotopic composition of the Sun:}
It is commonly assumed that the Sun has oxygen isotope ratios somewhere along the slope-1 line shown in Figure \ref{appendixfig1}.  Recently, this assumption was put to the test by mass spectrometric measurements of oxygen isotope ratios of solar wind returned by the Genesis mission \citep{McKeegan2010-, McKeegan2010V31B-2316}.  These preliminary results, after correction for mass fractionation in the concentrator where the samples were taken (based on $\rm ^{22}Ne/^{20}Ne$ fractionation), yield 
$\rm \delta^{18}O = -99 \pm 5 \permil (2\sigma)$ and  $\rm \delta^{17}O = -79 \pm 9 \permil$ relative to SMOW.  These values define an $\rm [^{18}O]/[^{17}O]$ of $5.07 \pm 0.08$ $(2\sigma)$, within error of the value for rocks as described above (Figure \ref{appendixfig1}).  We conclude that the $\rm [^{18}O]/[^{17}O]$ of the solar system as a whole, rather than just the rocky bodies, is well defined at $5.2 \pm 0.2$. 

\cite{Ayres2006-618} report oxygen isotope ratios for the Sun based on measurements of the abundances of CO isotopologues in the atmospheric layer above the photosphere cooled by rovibrational transitions in CO.  The absolute $\rm [^{16}O]/[^{18}O]$  and $\rm [^{16}O]/[^{17}O]$ ratios are sensitive to models for the thermal structure of the middle photosphere (see Table 7 in that work), but all of the derived ratios give $\rm [^{18}O]/[^{17}O]$ values closer to 4 than 5.  The preferred $\rm [^{16}O]/[^{18}O]$  and $\rm [^{16}O]/[^{17}O]$ ratios of $1700 \pm 220$ and $ 440 \pm 6$ correspond to a $\rm [^{18}O]/[^{17}O]$ of $3.9 \pm 0.5$ (uncertainties are based on standard errors about the means).  At face value, these results are in conflict with all mass spectrometric measurements of solar system materials, including those of the solar wind.  We offer no explanation for the disparity between oxygen isotope ratios determined for the COmosphere and those for the rest of the solar system.  In view of the larger uncertainties associated with the former, and the as yet poorly constrained systematic errors associated with solar atmosphere thermal models, we do not consider these solar spectroscopic measurements to be remotely as reliable as the mass spectrometric measurements.  Further study is warranted in all events.

\bibliographystyle{apj}

\newpage

\end{document}